\documentclass[amsmath,amssymb,twocolumn]{revtex4}

\usepackage{mathrsfs}
\usepackage{mathtools}
\usepackage{xfrac}
\usepackage{graphicx}
\usepackage{dcolumn}
\usepackage{bm}
\usepackage{color}
\usepackage{tabularx}
\usepackage{amsmath}

\newcommand{\be}{\begin{equation}}
\newcommand{\ee}{\end{equation}}

\newcommand{\beq}{\begin{eqnarray}}
\newcommand{\eeq}{\end{eqnarray}}
\newcommand{\bea}{\begin{align}}
\newcommand{\eea}{\end{align}}
\newcommand{\beqq}{\begin{eqnarray*}}
\newcommand{\eeqq}{\end{eqnarray*}}

\newcommand{\trace}[1]{\text{Tr}\,[#1 ]  }

\newcommand{\bra}[1]{\langle #1 | }
\newcommand{\ket}[1]{ | #1 \rangle }

\newcommand{\ve}{{\varepsilon}}

\begin{document}

\title{Identification of nematic superconductivity from the upper critical field}

\author{J\"orn W. F. Venderbos}
\affiliation{Department of Physics, Massachusetts Institute of Technology,
Cambridge, MA 02139, USA}

\author{Vladyslav Kozii}
\affiliation{Department of Physics, Massachusetts Institute of Technology,
Cambridge, MA 02139, USA}

\author{Liang Fu}
\affiliation{Department of Physics, Massachusetts Institute of Technology,
Cambridge, MA 02139, USA}

\begin{abstract}
Recent nuclear magnetic resonance and specific heat measurements have provided concurring evidence of spontaneously broken rotational symmetry in the superconducting state of the doped topological insulator Cu$_x$Bi$_2$Se$_3$. This suggests that the pairing symmetry corresponds to a two-dimensional representation of the $D_{3d}$ crystal point group, and that Cu$_x$Bi$_2$Se$_3$ is a nematic superconductor. In this work, we present a comprehensive study of the upper critical field $H_{c2}$ of nematic superconductors within Ginzburg-Landau (GL) theory. Contrary to typical GL theories which have an emergent U(1) rotational symmetry obscuring the discrete symmetry of the crystal, the theory of two-component superconductors in trigonal $D_{3d}$ crystals reflects the true crystal rotation symmetry. This has direct implications for the upper critical field. 
First, $H_{c2}$ of trigonal superconductors with $D_{3d}$ symmetry exhibits a sixfold anisotropy in the basal plane. Second, when the degeneracy of the two components is lifted by, e.g., uniaxial strain, $H_{c2}$ exhibits a twofold anisotropy with characteristic angle and temperature dependence. Our thorough study shows that measurement of the upper critical field is a direct method of detecting nematic superconductivity, which is directly applicable to  recently-discovered trigonal superconductors Cu$_x$Bi$_2$Se$_3$, Sr$_x$Bi$_2$Se$_3$, Nb$_x$Bi$_2$Se$_3$, and Tl$_x$Bi$_2$Te$_3$. 
\end{abstract}

\maketitle

\section{Introduction}

Unconventional superconductors can be defined by superconducting order parameters that transform nontrivially under crystal symmetries. 
For a given superconductor, possible unconventional order parameters are classified by non-identity representations of the crystal point group.  Such representations are either one-dimensional or multi-dimensional, and this distinction defines two classes of unconventional superconductivity \cite{sigrist91,mineevbook}. The first class is exemplified by $d$-wave superconductors in cuprates \cite{harlingen95,tsuei00}, while the second class is exemplified by the $p$-wave superconductivity in Sr$_2$RuO$_4$ \cite{mackenzie03},  with two degenerate components $(p_x, p_y)$ at the superconducting transition temperature.   
Superconducting states in the second class spontaneously break lattice or time-reversal symmetry~\cite{volovik85}, in addition to the U(1) gauge symmetry, leading to novel thermodynamic and transport properties not seen in single-component superconductors. The search for new superconductors with multi-component order parameters is therefore of great interest.    

The doped topological insulator Cu$_x$Bi$_2$Se$_3$, a superconductor with $T_c \sim 3.8$K \cite{cava, ando}, has recently attracted a lot of attention as a promising candidate for unconventional superconductivity~\cite{fuberg,hao11,bay12,nagai12,lawson12,hashimoto13,yip13,wan14,brydon14,schneeloch15,andofu}. Fu and Berg proposed that it may have an odd-parity pairing symmetry resulting from inter-orbital pairing in a strongly spin-orbit-coupled normal state \cite{fuberg}. While previous surface-sensitive experiments \cite{point-contact, stm} drew disparate conclusions regarding the nature of superconductivity in this material, direct tests of the pairing symmetry in the {\it bulk} of Cu$_x$Bi$_2$Se$_3$ are carried out only very recently. 
A nuclear magnetic resonance (NMR) measurement \cite{Zheng} found that despite the three-fold rotational symmetry of the crystal, the Knight shift displays a twofold anisotropy below $T_c$  as the field is rotated in the basal plane. The twofold anisotropy is also found in the specific heat of the superconducting state under magnetic fields down to $H=0.03$T corresponding to $H/H_{c2} \sim 0.015$ \cite{maeno}. Both experiments found that the twofold anisotropy vanishes in the normal state, establishing that the superconducting state of Cu$_x$Bi$_2$Se$_3$ spontaneously breaks the three-fold rotational symmetry. This is only possible when the order parameter belongs to the two-dimensional $E_u$ or $E_g$ representation of the $D_{3d}$ point group.  The $E_g$ pairing has been ruled out by comparing the theoretically expected gap structure \cite{fu14} with specific heat data \cite{ando, maeno}.   
These results taken together strongly suggest that the pairing symmetry of Cu$_x$Bi$_2$Se$_3$ is $E_u$, an odd-parity pairing with two-component order parameters \cite{fuberg}. 

Spontaneous rotational symmetry breaking due to superconductivity is a rare and remarkable phenomenon. Superconductors exhibiting rotational symmetry breaking from multi-component order parameters can be called nematic superconductors \cite{fu14}, in analogy with the nematic liquid crystals and nematic electronic states in non-superconducting metals \cite{kivelson98,fradkin}.     
Nematic and chiral superconductivity, the latter breaking time-reversal symmetry, are the two distinct and competing states of multi-component superconductors, corresponding to real and complex order parameters respectively \cite{volovik85,sigrist91}. Broken rotational symmetry has previously been reported in the heavy-fermion superconductor UPt$_3$ \cite{joynt02} under a magnetic field \cite{machida12}. In addition, the A phase in a narrow temperature range at zero field is likely rotational symmetry breaking, which however may be due to antiferromagnetic order already present in the normal state \cite{aeppli88,fisher89}. Thus the recent discovery of broken rotational symmetry in Cu$_x$Bi$_2$Se$_3$, without broken time-reversal symmetry, may potentially open a fruitful research direction.

Motivated by the recent experimental progress, in this work we study the upper critical field $H_{c2}$ of trigonal nematic superconductors within the framework of Ginzburg-Landau (GL) theory. Such GL theory admits a new trigonal gradient term which is not allowed in hexagonal crystals~\cite{mineev}. We relate the gradient terms to Fermi surface and gap function anisotropies by a microscopic calculation of the GL coefficients. Building on and generalizing the previous work~\cite{mineev}, we show that the upper critical field generically displays a sixfold anisotropy within the basal plane of trigonal crystals. We further show that a uniaxial strain acts as a symmetry-breaking field in nematic superconductors, which directly couples to the bilinear of the two-component superconducting order parameter. As a result, $H_{c2}$ in the basal plane exhibits a twofold anisotropy with a distinctive angle and temperature dependence, similar to theoretically expected results for UPt$_3$ in the presence of anti-ferromagnetic order~\cite{agterberg94,agterberg95}. 
Our findings suggest that measurement of the upper critical field is a direct method of detecting nematic superconductivity. In particular, this method may shed light on the pairing symmetries of other superconducting doped topological insulators Sr$_x$Bi$_2$Se$_3$ \cite{sr-doped,sr-doped-2}, Nb$_x$Bi$_2$Se$_3$ \cite{nb-doped} and Tl$_x$Bi$_2$Te$_3$ \cite{tl-doped}, which have yet to be determined.

\section{Ginzburg-Landau theory} 

We start by constructing the GL theory of odd-parity two-component superconductivity in crystals with $D_{3d}$ point group and strong spin-orbit coupling. The pairing potential $\hat{\Delta}(\vec{k})$, which is a $\vec k$-dependent matrix in spin space, takes the following form
\begin{gather}
\hat{\Delta}(\vec{k}) = \eta_1 \hat{\Delta}_1(\vec{k})+\eta_2 \hat{\Delta}_2(\vec{k}). \label{eq:pairpotential}
\end{gather}
The pairing potential is a linear superposition of two degenerate components $\hat{\Delta}_{1,2}(\vec{k})$, the basis functions of the two-dimensional pairing channel $E_u$ (specific gap functions are given in the Supplementary Material, Sec. III). For odd-parity superconductors the pairing components satisfy $\hat{\Delta}_{1,2}(-\vec{k}) = -\hat{\Delta}_{1,2}(\vec{k}) $. As basis functions of $E_u$, the two partners $\hat{\Delta}_{1,2}(\vec{k})$ transform differently under the mirror symmetry $x \to -x$, i.e., $\hat{\Delta}_{1}(\vec{k})$ is even whereas $\hat{\Delta}_{2}(\vec{k})$ is odd. A key property of (doped) Bi$_2$Se$_3$ materials is strong spin-orbit coupling that locks the electron spin to the lattice. The two complex fields $\eta_{1,2}$ define the superconducting order parameters $\eta = (\eta_1,\eta_2)^T$. In contrast, in case of triplet superconductors in spin-rotation invariant materials the order parameter components are vectors in spin space.

The GL theory of  two-component superconductivity is formulated in terms of the order parameters $\eta$ and the GL free energy $F_{\text{tot}}= \int d^3\vec{x} \, f_{\text{tot}}$ is the sum of a homogeneous term and a gradient term given by $f_{\text{tot}}   =  f_{\text{hom}} + f_{D}  $, where $f_{\text{hom}}$ and $f_{D}$ are the corresponding free energy densities.
In addition, the free energy contains a Maxwell term $f_{\text{EM}} = (\vec{\partial}\times \vec{A})^2/8\pi$, which for our purposes can be taken as a constant. The free energy densities $f_{\text{hom}}$ and $f_{D}$ are polynomial expansions in the order parameter fields and their gradients, and consist of all terms invariant under the symmetry group of the crystal. For two-component trigonal superconductors the homogeneous contribution is the same as the corresponding expression for hexagonal symmetry~\cite{volovik85,sigrist91}, 
\begin{gather}
f_{\text{hom}}  = A\eta^\dagger\eta + B_1(\eta^\dagger\eta)^2 + B_2|\eta^*_1\eta_2-\eta^*_2\eta_1|^2, \label{eq:Fhom}
\end{gather}
to fourth order in $\eta$, and we have defined $\eta^\dagger = (\eta^*_1, \eta^*_2)$. The coefficients $A \propto T-T_c$ and $B_{1,2}$ are phenomenological constants of the GL theory. The sign of GL coefficient $B_2$ determines the nature of the superconducting state, selecting either chiral or nematic order \cite{fu14, Venderbos}.

Spatial variation of the superconducting order parameter is captured by the gauge-invariant gradient $D_i   = -i\partial_i -qA_i$, with $\vec{A}$ the electromagnetic vector potential and $q=-2e$. In case of multicomponent order parameters, there generally exist multiple independent gradient terms which are allowed by crystal symmetry. It is insightful to present all gradient terms in order of ``emergent symmetry''.  For crystals with a principal rotation axis along the $z$ direction, such as the three- and sixfold rotations of trigonal and hexagonal crystals, four gradient terms with full continuous in-plane rotational symmetry are present and given by~\cite{sigrist91,sauls94,zhitomirsky95}
\begin{multline}
 f_{D} =  J_1 (D_i \eta_a)^*D_i \eta_a + J_2 \epsilon_{ij}\epsilon_{ab}(D_i \eta_a)^*D_j \eta_b   \\
+J_3 (D_z \eta_a)^*D_z \eta_a  +J_4\left[ |D_x \eta_1|^2 + |D_y \eta_2|^2  \right.  \\ 
- |D_x \eta_2|^2 - |D_y \eta_1|^2  +  (D_x \eta_1)^*D_y \eta_2  + (D_y \eta_1)^*D_x \eta_2  \\
\left.   +  (D_x \eta_2)^*D_y \eta_1 + (D_y \eta_2)^*D_x \eta_1   \right]   \label{eq:stiffhexa} 
\end{multline}
 (summation understood, $i=x,y$, $a=1,2$), and  $J_{1,2,3,4}$ are the phenomenological GL coefficients. The first three terms are invariant under independent U(1) rotation of coordinates and order parameters, and thus have an emergent U(1)$\times$U(1) symmetry, whereas
the gradient term with coefficient $J_4$ is invariant under arbitrary joint rotations of coordinates and order parameters, i.e., an emergent U(1) symmetry. Therefore, $f_D$ does not reflect the discrete rotational symmetry of the crystal. However, a new gradient term $f_{D,\text{trig}}$, which we call trigonal gradient term, is uniquely present in crystals with trigonal symmetry, but not allowed in hexagonal crystals~\cite{mineev}. It is given by the expression
\begin{multline}
 f_{D,\text{trig}} = J_5 \left[ (D_z \eta_1)^*D_x \eta_2  + (D_z \eta_2)^*D_x \eta_1 \right.  \\
\left. + (D_z \eta_1)^*D_y \eta_1 - (D_z \eta_2)^*D_y \eta_2+ c.c. \right]. \label{eq:stifftrig}
\end{multline}
The appearance of this new gradient term, which has $D_{3d}$ symmetry, can be understood from angular momentum, since in trigonal symmetry $L=3$ is equivalent to $L=0$. Indeed, in momentum space ($D_i \rightarrow q_i$) the trigonal gradient term can be expressed as $iq_z( q_-\eta^*_+\eta_- -q_+\eta^*_-\eta_+ )$, where $q_\pm = q_x \pm i q_y$ and similarly for $\eta_{1,2}$. The relative phases between $\eta_+$ ($q_+$) and $\eta_-$ ($q_-$) are determined by 
mirror symmetry: $\eta_1$ ($\eta_2$) is even (odd) under $x \rightarrow -x $. It follows from the structure of $f_{D,\text{trig}}$ that the spatial variation of the order parameter in the basal plane is coupled to spatial variation in the $z$-direction, which is in sharp contrast to hexagonal and tetragonal crystals. In the rest of this work we map out the consequences of trigonal crystal anisotropy in the GL theory for the upper critical field.

\section{Upper critical field in the basal plane} 

The angular dependence of $H_{c2}$ was first proposed as a method to establish the multicomponent nature of unconventional superconductors in the context of heavy-fermion superconductors \cite{gorkov84,burlachkov85,machida85}. The key idea is as follows. For the class of single-component (e.g., $s$-wave) superconductors with trigonal, tetragonal, and hexagonal symmetry, $H_{c2}$ is always isotropic within the GL theory, due to the emergence of $U(1)$ rotational symmetry to second order in the gradients. In case of multicomponent superconductors, effects of crystal anisotropy can appear in the GL theory, removing the emergent U(1) symmetry, but this crucially depends on crystal symmetry. For instance, hexagonal systems with multicomponent order parameters do not show in-plane $H_{c2}$-anisotropy due to the emergent rotational symmetry of Eq.~\eqref{eq:stiffhexa}, whereas tetragonal symmetry can give rise to an angular dependence of $H_{c2}$ with fourfold symmetry~\cite{burlachkov85}. In trigonal crystals, $H_{c2}$ can exhibit a sixfold anisotropy in the basal plane~\cite{mineev} as of Eq.~\eqref{eq:stifftrig}. Here we map out the basal plane upper critical field of trigonal superconductors for general GL gradient coefficients.

Within GL theory, the upper critical field is calculated by solving the GL equations obtained from $F_{\text{tot}}$, keeping only terms linear in $\eta$ since the order parameter is small at $H_{c2}$. Therefore, the calculation also applies to chiral superconductors. The resulting system of GL equations, which is given by
\begin{multline}
-A\eta_a  = J_1 (D_x^2 + D_y^2 )\eta_a + J_3 D_z^2 \eta_a + J_2 \epsilon_{ab}[D_x,D_y]\eta_b  \\
+ J_4\left[ (D^2_x-D^2_y)\tau^z_{ab}+ \{D_x,D_y\} \tau^x_{ab} \right]\eta_b \\
+J_5\left[  \{D_z,D_x\} \tau^x_{ab} + \{D_z,D_y\} \tau^z_{ab} \right]\eta_b, \label{eq:GLeq}
\end{multline}
can be solved as a two-component harmonic oscillator problem, leading to a Landau-level spectrum from which $H_{c2}$ is determined as the lowest Landau-level solution. The coupling of the two harmonic oscillators is determined by the structure of the GL equations, and is in general complicated by the presence of multiple gradient terms. In hexagonal and tetragonal systems, straightforward or even exact analytical expressions for $H_{c2}$ can be found~\cite{burlachkov85}. In contrast, the trigonal gradient term of Eq.~\eqref{eq:stifftrig} couples basal plane gradients to gradients in the orthogonal direction, giving rise to a different set of harmonic oscillator equations to which previous methods do not apply. A special limiting case was considered in Ref.~\onlinecite{mineev}. We generalize this result by solving the GL equations in the presence an in-plane magnetic field for general gradient coefficients. In deriving the general solution we adopt an operator based approach and exploit that harmonic oscillator mode operators corresponding different cyclotron frequencies can be related by squeezing operators. Here we present and discuss the main results, and give a detailed account of the lengthy calculations in the Supplemental Material (SM). For convenience, below we will refer to the appropriate section of the SM.

To demonstrate the key features of $H_{c2}$ in trigonal crystals, we will focus the discussion on the most physical case, where trigonal anisotropy effects may be considered weak and $J_5$ can be treated as perturbation. We take the magnetic field $ \vec{H}$ in the basal plane to be given by $\vec{H} = H (\cos\theta,\sin\theta,0)^T$, which corresponds to a vector potential $\vec{A} = Hz (\sin\theta, -\cos\theta,0 )^T$. It is convenient to rotate the basal plane GL gradients $D_{x,y} = -i \partial_{x,y} +2eA_{x,y}$ according to the transformation 
\begin{gather}
 \begin{pmatrix} D_{\parallel} \\  D_{\perp} \end{pmatrix}  =  \begin{pmatrix} \cos\theta & \sin\theta \\ \sin\theta & -\cos\theta \end{pmatrix}   \begin{pmatrix} D_x \\  D_y \end{pmatrix} , \label{eq:Dtransform}
\end{gather}
such that $D_{\parallel}$ is along the field and $D_{\perp}$ is perpendicular to the field. These operators satisfy $[D_{\parallel},  D_{\perp} ] = [D_{\parallel},  D_z ] =0$, and $ D_{\perp}$ and $D_z$ define the magnetic algebra $[ D_z, D_{\perp} ] =-2ie H$. Writing Eq. \eqref{eq:GLeq} in terms of $ D_{\perp}$ and $D_z$,  and setting $D_{\parallel} \eta_a = 0$ (i.e., no modulation along the field), one obtains
\begin{multline}
-A  \eta_a =  (J_1D^2_{\perp} + J_3 D^2_z)\eta_a \\
- J_4D^2_{\perp} ( \cos 2\theta \tau^z_{ab}+ \sin 2\theta \tau^x_{ab})\eta_{b} \\
+  J_5 \{D_z,D_{\perp}\} ( -\cos \theta \tau^z_{ab}+ \sin \theta \tau^x_{ab})\eta_{b}.  \label{eq:trigsystem}
\end{multline}
Next, it is convenient to diagonalize the term proportional to $J_4$. This is achieved by a the rotation of the order parameters given by
\begin{gather}
 \begin{pmatrix} \eta_1 \\  \eta_2 \end{pmatrix}  =  \begin{pmatrix} \cos\theta & - \sin\theta \\ \sin\theta & \cos\theta \end{pmatrix}   \begin{pmatrix} f_1 \\  f_2 \end{pmatrix}.  \label{eq:etatransform}
\end{gather}
In terms of the rotated order parameters $(f_1,f_2)^T$ the GL equations read
\begin{widetext}
\begin{gather}
-A  \begin{pmatrix} f_1 \\  f_2 \end{pmatrix} = \begin{pmatrix} J_3D^2_z +(J_1-J_4)D^2_{\perp} & 0 \\ 0 &J_3D^2_z +(J_1+J_4)D^2_{\perp}\end{pmatrix}\begin{pmatrix} f_1 \\  f_2 \end{pmatrix}  +  J_5 \{D_z,D_{\perp}\} \begin{pmatrix} -\cos 3\theta &  \sin 3\theta \\ \sin 3\theta & \cos 3\theta \end{pmatrix} \begin{pmatrix} f_1 \\  f_2 \end{pmatrix}    \label{eq:trigonalrotated}.
\end{gather}
\end{widetext}
Note that only the term proportional to $J_5$ depends on the angle $\theta$. We now describe solutions to Eq. \eqref{eq:trigonalrotated} obtained by treating $J_5$ as a perturbation. 

To start, let us consider taking both $J_4=J_5=0$. Solving the GL equations then yields two degenerate series of Landau levels with cyclotron frequency $\omega = \sqrt{J_1J_3}$, with the upper critical field given by $H_{c2}  = - A /2e\omega = - A /2e\sqrt{J_1J_3}$ (more details are provided in Sec. II B of the SM). Including the gradient contribution in Eq.~\eqref{eq:stiffhexa} proportional to $J_4$ simply makes the cyclotron frequencies inequivalent, $\omega_{1,2} = \omega\sqrt{1\mp |J_4|J_3/\omega^2 }= \omega\sqrt{1\mp |J_4|/J_1 }$, and increases the upper critical field to $H_{c2}  = - A /2e\omega_1$. This defines the exactly solvable unperturbed system. Then, introducing trigonal perturbation parametrized by $J_5$ couples the two series of Landau levels with different frequencies in a nontrivial way: the coupling of in-plane and out-of-plane gradients implies a coupling of canonically conjugate operators of the form $\{D_z, D_{\perp} \} \sim \{-i\partial_z, z \}$. To solve the system of GL equations we assume that crystal anisotropy effects are weak and use second order perturbation theory to obtain the correction to the cyclotron frequency $-\delta\omega_1$. (The calculations are lengthy and described in detail in Sec. II B 3 of the SM.) The upper critical field then becomes $H_{c2}  = \widetilde{H}_{c2} ( 1 + \delta\omega_1/\omega_1)$ with $\widetilde{H}_{c2} \equiv - A /2e\omega_1$. We find $H_{c2}$ to lowest order in $J_5$ as
\begin{gather}
\frac{H_{c2}(\theta)}{\widetilde{H}_{c2}} = 1 + \frac{J^2_5}{2\omega^2_+}\left[\frac{\cos^2 3\theta}{(1-\tfrac{\omega_-}{\omega_+})^2} + \frac{\sin^2 3\theta }{1-\tfrac{\omega_-}{\omega_+}} F(\tfrac{\omega_-}{\omega_+}) \right] ,  \label{eq:Hc2trig}
\end{gather}
where the frequencies $\omega_\pm$ are defined as $\omega_\pm = (\omega_2\pm \omega_1)/2$. In the limit of small $J_4/J_1$ these frequencies become $\omega_+ \sim \omega$ and $\omega_- \sim \omega |J_4|/2J_1$. The function $F(x)$ arises due to the coupling of two series of Landau levels with different cyclotron frequencies and oscillator eigenfunctions. It takes the form
\begin{eqnarray}
F(x) &=&  \frac{ \left(1- x^2 \right)^{\tfrac{5}{2}}}{x^2}\sum^{\infty}_{m \ge 0} \frac{(2m)!}{(m!)^2 4^m} \frac{x^{2m}(2m- \frac{x^2}{1-x^2} )^2}{2m+ x(2m +1)} \nonumber \\
&=&\frac{1-x}x \left[ \sqrt{\frac{1+x}{1-x}}\, {}_{2}F_1 \left( \tfrac12, \tfrac{a}{2}; 1+\tfrac{a}{2};x^2  \right) -1\right]
\end{eqnarray}
where $a = x/(1+x)$ and ${}_{2}F_1[\alpha,\beta;\delta;\gamma]$ is a hypergeometric function. The function $F(x)$ has the property $F(0)=1$, which implies that for $J_4=0$ (corresponding to $\omega_-/\omega_+=0$) no angular dependence of $H_{c2}$ exists. The latter is a consequence of an emergent rotational symmetry of $f_{D,\text{trig}}$ in Eq.~\eqref{eq:stifftrig}: it is invariant under in-plane rotations of the order parameters and coordinates according to $q_+ \rightarrow q_+e^{2i\varphi}$, $\eta_+ \rightarrow \eta_+e^{-i\varphi}$. (Note that this is not a physical symmetry.) 

In general, however, considering all regimes of gradient coefficients that satisfy the stability constraints of the free energy, $H_{c2}$ exhibits a six-fold anisotropy in the basal plane of the crystal. For instance, the sixfold $H_{c2}$-anisotropy can be obtained starting from a solution of the GL equations derived from Eqs.~\eqref{eq:stiffhexa} and \eqref{eq:stifftrig} for $J_5\neq 0$ and $J_4=0$, and treating $J_4$ as a small perturbation. This case was considered in Ref.~\onlinecite{mineev} and is described in Sec. II B 2 of the SM.

\begin{figure}
\includegraphics[width=\columnwidth]{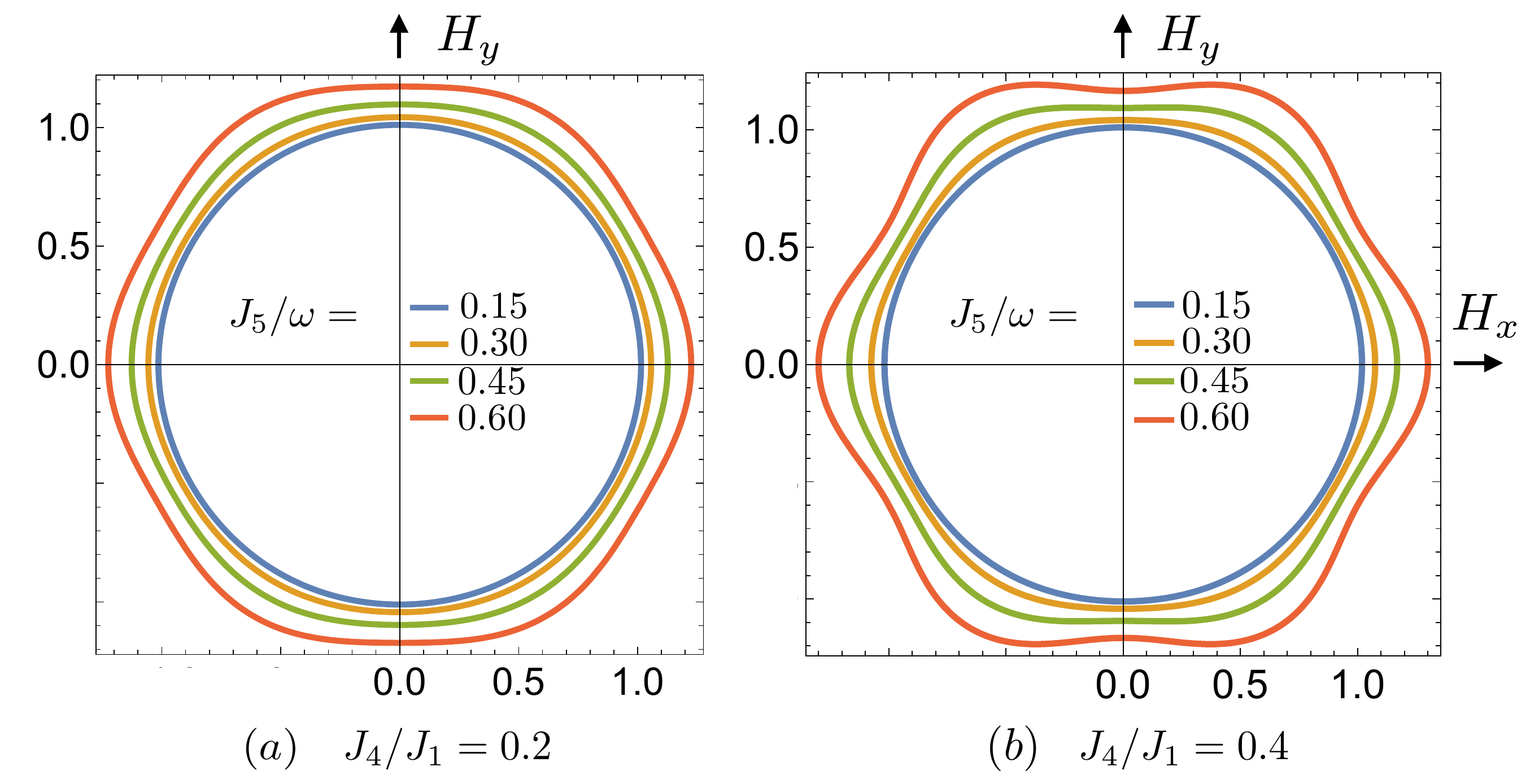}
\caption{\label{fig:Hc2_6fold} Upper critical field ($H_{c2}$) anisotropy of two-component pairing in trigonal crystals with $D_{3d}$ point group symmetry, originating from the trigonal GL anisotropy term \eqref{eq:stifftrig}. (a) Polar plot of the angular dependence of $H_{c2}$ with six-fold symmetry given by Eq.~\eqref{eq:Hc2trig} (normalized by $\widetilde{H}_{c2}$) for $J_4/J_1=0.2$. Different curves correspond to $J_5/\omega= J_5/\sqrt{J_1J_3} = (0.15,0.30,0.45,0.60)$ (inward to outward). (b) Same as (a) but for $J_4/J_1=0.4$. 
 }
\end{figure}

Figure~\ref{fig:Hc2_6fold} shows the angular dependence of the upper critical field for small to moderate $J_5/\omega = J_5/\sqrt{J_1J_3}$ and $J_4/J_1$ as obtained from Eq.~\eqref{eq:Hc2trig}. Note that in general, for materials with weak to moderate (crystal) anisotropy effects, one expects $J_1\sim J_3$. To make the interplay between $J_4$ and $J_5$ explicit, we expand Eq.~\eqref{eq:Hc2trig} for small $J_4/J_1$ and find
\begin{gather}
\frac{H_{c2}(\theta)}{H_{c2}(\tfrac{\pi}{4})} = 1 + h\cos 6\theta ,  \label{eq:expand}
\end{gather}
where  $h = 3|J_4|J^2_5/16J^2_1J_3$. This expression serves to highlight an important feature of the angular dependence of $H_{c2}$: $H_{c2}(\theta=\pi/2)/H_{c2}(\theta=0) < 1$, which is independent of system specific parameters. Here $\theta=0$ is defined by an axis orthogonal to a mirror plane. 

Within weak coupling, the GL coefficients $J_i$ can be obtained in terms of Fermi surface and gap function properties using a microscopic mean-field Hamiltonian with pairing potential $\hat{\Delta}(\vec{k}) $ given by Eq.~\eqref{eq:pairpotential}. The gradient coefficients $J_1$, $J_3$, $J_4$, and $J_5$ are proportional to $N(\varepsilon_F)v^2_F/T^2_c \sim N(\varepsilon_F)\xi^2_0$, where $\varepsilon_F$,  $v_F$, and $\xi_0$ are the Fermi energy, Fermi velocity, and correlation length respectively, and $N(\varepsilon_F)$ is the density of states. (The calculations are presented in detail in Sec. III of the SM.) We find that their relative strength depends on the crystal anisotropy of the Fermi surface and of the gap functions $\hat{\Delta}_{1,2}(\vec{k})$.  In particular, $J_5$ is nonzero only when trigonal Fermi surface anisotropy is present, or when the gap function is composed of trigonal crystal spherical harmonics of the $E_u$ pairing channel (see Sec. III A of the SM), and is generally expected to be weak. 

The general sixfold basal plane anisotropy of $H_{c2}$ is a direct consequence of trigonal symmetry and a discriminating characteristic of two-component pairing symmetry. Indeed, single-component superconductivity corresponding to one-dimensional pairing channels of point group $D_{3d}$ cannot exhibit sixfold $H_{c2}$ anisotropy: the in-plane gradient term is given by $\tilde{J}_1 |D_i \psi|^2$ and has emergent U(1) rotational symmetry.  As a result, the sixfold anisotropy provides a clear experimental evidence for two-component pairing.

\section{Nematic superconductivity and upper critical field} 

Within our GL theory, the rotational symmetry breaking superconducting state reported in Refs.~\cite{Zheng,maeno} corresponds to a \emph{real} order parameter, i.e., $\eta = \eta_0 (\cos \phi, \sin \phi )^T$. Up to fourth order [see Eq.~\eqref{eq:Fhom}], the angle $\phi$ represents a continuous degeneracy. This degeneracy is lifted at sixth order by a crystal anisotropy term and leads to a discrete set of degenerate ground states \cite{fu14, Venderbos}. In materials, such as Cu$_x$Bi$_2$Se$_3$, the remaining degeneracy may be further lifted by a symmetry-breaking pinning field, selecting a unique ground state. The origin of such pinning can be strain-induced distortions of the crystal \cite{hicks14}, but in principle, any order with the same symmetry, electronic or structural, can pin the order parameter. In case of two-component superconductors, the symmetry-breaking (SB) pinning field couples \emph{linearly} to order parameter $\eta$ in the following way
\begin{gather}
 f_{\text{SB}} =    g \left[(u_{xx} - u_{yy})(|\eta_1|^2 - |\eta_2|^2)  + 2  u_{xy} (\eta_1^*\eta_2 + \eta_2^*\eta_1) \right], \label{eq:SBfield}
\end{gather}
with coupling constant $g$. The order parameter bilinears $(|\eta_1|^2 - |\eta_2|^2,\eta_1^*\eta_2 + \eta_2^*\eta_1)$ constitute a two-component subsidiary nematic order parameter~\cite{fu14} with the same symmetry as the symmetry-breaking field $(u_{xx} - u_{yy},2  u_{xy})$. For comparison, uniaxial strain in single-component superconductors couples to the gradient of the order parameter $\psi$, taking the form $\tilde{J}_{1,x} |D_x \psi |^2+\tilde{J}_{2,y} |D_y \psi |^2$ different from Eq.~\eqref{eq:SBfield}. It is worth noting that the coupling considered here differs from the candidate theories proposed for the hexagonal superconductor UPt$_3$, in which case magnetic order couples quadratically, instead of linearly, to order parameter bilinears ~\cite{aeppli88,hess89,machida89,sauls94,zhitomirsky95,joynt02}.  

From a microscopic perspective, the origin of the order parameter pinning in Eq.~\eqref{eq:SBfield} can be understood as a (strain-induced) Fermi surface distortion, leading to different Fermi velocities $v_{F,x}\neq v_{F,y}$. A uniaxial distortion of this form couples to $|\eta_1|^2 - |\eta_2|^2$ and has the effect of selecting either $\eta=(1,0)$ or $\eta=(0,1)$ by raising $T_c$, resulting in a split transition. A quantitative calculation of the coupling constant $g$, relating the order parameter bilinear to such Fermi surface distortion can be obtained within weak-coupling (see \cite{suppmat}). This effect of a Fermi surface distortion should be compared to uniaxial gradient anisotropies such as $\sim |D_x \eta _a|^2- |D_y \eta_a |^2$ and $\sim |D_i \eta _1|^2- |D_i \eta_2 |^2$, with the effect of the former being enhanced by a factor of $\ln (\omega_D /T_c) (\xi/\xi_0)^2$~\cite{suppmat}, where $\xi$ is the coherence length, $\ln \omega_D /T_c \sim 1/V N(\varepsilon_F)$, $\omega_D$ is a cutoff frequency, and $V$ is an effective interaction energy scale associated with the pairing. In addition, the effect of a uniaxial Fermi surface distortion $\sim v_{F,x}/ v_{F,y}$ on the shift of $T_c$ is enhanced by $\ln \omega_D /T_c $.

To address the effect of the SB field on $H_{c2}$ in case of the trigonal nematic superconductors, we solve the linearized GL equations for small $J_{4,5}$ gradient coefficients in the presence of a uniaxial symmetry breaking term defined as $\delta (|\eta_1|^2 - |\eta_2|^2)$, taking $\delta$ as a measure of the uniaxial anisotropy. Here we focus the discussion on the most salient features, for which we take $J_5=0$, and relegate a more detailed account to the SM. A similar problem of upper critical field anisotropy was studied for split transitions in UPt$_3$ \cite{agterberg94,agterberg95}.

Setting $J_5=0$ in Eq. \eqref{eq:trigsystem} and adding the contribution from the symmetry breaking field, the GL equations take the form
\begin{multline}
-A  \eta_a =  (J_1D^2_{\perp} + J_3 D^2_z)\eta_a  +  \delta\tau^z_{ab} \eta_b \\  
- J_4D^2_{\perp} ( \cos 2\theta \tau^z_{ab}+ \sin 2\theta \tau^x_{ab})\eta_b\label{eq:GLsymbreak}.
\end{multline}
The upper critical field is obtained by using the magnetic algebra of $D_z$ and $D_{\perp}$, and projecting into the lowest Landau level. The upper critical field is then determined from the following implicit equation (see Sec. II D of the SM)
\be
\frac{-A}{\omega}  =  \frac{1}{l^2_b}  - \sqrt{ \frac{J^2_4J^2_3}{4\omega^4 l^4_b} + \frac{\delta ^2}{\omega^2}-  \frac{J_4J_3\delta  }{\omega^3 l^2_b}\cos 2\theta  }, \label{eq:Hc2SB}
\ee
(recall $\omega = \sqrt{J_1J_3}$) where the magnetic length $l_b$ is defined as $2eH=1/l^2_b$. For $\delta=0$ we recover the result for $J_5=0$ in Eq.~\eqref{eq:Hc2trig}, to first order in $J_4/J_1$ (i.e., $\omega_1$ expanded to first order in $J_4/J_1$). For $J_4=0$ we simply find $H_{c2} = H_{c2,0}$ [see Eq.~\eqref{eq:expand}], but with critical temperature $T^*_c = T_c + \Delta T_c$ with $\Delta T_c \sim |\delta|$. This follows from comparing $\delta $ to $A \sim (T - T_c)$, i.e., $\delta$ shifts the transition temperature and can be taken as a measure of $T$. We define a dimensionless temperature $t$ by $T = T^*_c - t \Delta T_c$. 

\begin{figure}
\includegraphics[width=\columnwidth]{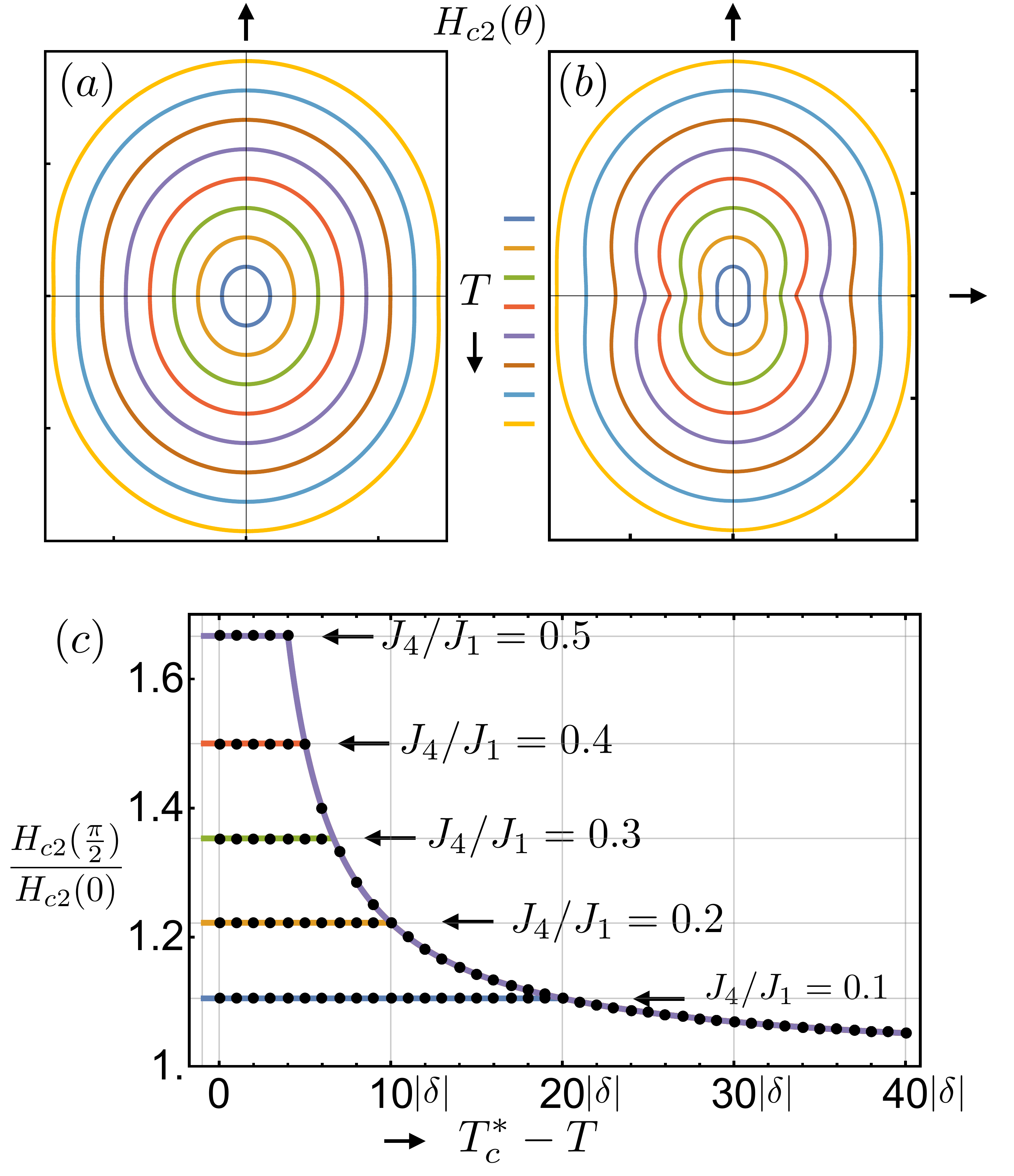}
\caption{\label{fig:Hc2_2fold} (a) Polar plot of the angular dependence of $H_{c2}$ in the presence of a symmetry-breaking field $\delta$ for $J_4/J_1 = 0.1$, calculated using Eq.~\eqref{eq:Hc2SB} (in arbitrary units of $H$). Different curves represent different temperatures: $T= T^*_c - t \Delta T_c$ (recall that $\Delta T_c \sim |\delta|$), where $t=1,\ldots,8$ and the outermost curve corresponds to $t=8$.  (b) Same as in (a) but for relatively large $J_4/J_1 = 0.6$. Figure (b) clearly shows the two-fold ``peanut''-shape anisotropy expected for two-component superconductors in the presence of a symmetry breaking field. (c) Plot of the $H_{c2}$-anisotropy coefficient $H_{c2}(\tfrac{\pi}{2})/H_{c2}(0)$ as function of effective temperature $t$ for various values of $J_4/J_1$. The horizontal grid lines correspond to the values $(1 + J_4/2J_1)/(1 - J_4/2J_1)$.
}
\end{figure}

For general $J_4/J_1$ and nonzero $\delta$ we solve Eq.~\eqref{eq:Hc2SB} for $H_{c2}$ and show the representative results for $J_4/J_1=0.1$ and $J_4/J_1=0.6$ in Figs.~\ref{fig:Hc2_2fold}(a) and \ref{fig:Hc2_2fold}(b). Two key characteristics of $H_{c2}$ in the presence of a pinning field are evident in Fig.~\ref{fig:Hc2_2fold}(a)--(b). First, the angular dependence of $H_{c2}$ exhibits a distinct two-fold anisotropy, with a typical ``peanut''-shape close to $T^*_c$. This twofold anisotropy becomes more pronounced with increasing $J_4/J_1$, as shown Fig.~\ref{fig:Hc2_2fold}(b). Expanding the square root in Eq.~\eqref{eq:Hc2SB} under the assumption of very small fields, i.e., $l^2_b \gg J_4J_3/2\omega \delta$, one finds $H_{c2} \propto (1 - J_4\text{sgn}(\delta) \cos 2\theta /2J_1)$ (see Sec. II D of the SM). This ``peanut''-shape of the $H_{c2}$ profile should be contrasted with the $H_{c2}$ profile of single-component superconductor 
where uniaxial gradient anisotropy leads to a weak \emph{elliptical} angular dependence of $H_{c2}$, an effect which is parametrically smaller than the twofold anisotropy in the two-component case. Consequently, the twofold anisotropy of $H_{c2}$ shown in Fig.~\ref{fig:Hc2_2fold}, in particular the ``peanut''-shape, is a discriminating property of two-component pairing. 

Second, the angular dependence of $H_{c2}$ is a function of temperature and has different shape in the vicinity of $T^*_c$ (i.e., small fields) as compared to far below $T_c$ (and high fields). This is in sharp contrast to the usual case, for instance Eq.~\eqref{eq:Hc2trig}, where only the overall magnitude of $H_{c2}$ is temperature dependent. The unusual temperature dependence of $H_{c2}$ can be more precisely captured by considering the upper critical field anisotropy ratio $H_{c2}(\tfrac{\pi}{2})/H_{c2}(0)$ as function of temperature. In the vicinity of $T^*_c$, the anisotropy ratio should exhibit temperature independent behavior given by $\sim (1 + J_4\text{sgn}(\delta)/2J_1)/(1 - J_4\text{sgn}(\delta)/2J_1)$ (see Sec. II D of the SM). This is shown in Fig.~\ref{fig:Hc2_2fold}(c), where the $H_{c2}$-anisotropy ratio is plotted for various values of $J_4/J_1$. In contrast, using Eq.~\eqref{eq:Hc2SB} we find that the $H_{c2}$-anisotropy ratio approaches unity for large temperature $t$ according to $\sim 2/(t-1)$, which is independent of GL parameters. Within the model of Eq.~\eqref{eq:Hc2SB}, the temperature at which the transition between two behaviors occurs is given by $t = 2J_1/|J_4|$. This ``kink'' feature was also found and discussed in the context of a hexagonal applicable to UPt$_3$~\cite{agterberg94,agterberg95,sauls96}. The distinctive temperature dependence of $H_{c2}$-anisotropy is uniquely associated with two-component pairing since single-component pairing with uniaxial gradient anisotropy leads to temperature independent $H_{c2}$-anisotropy.

\section{Discussion and conclusion}

To summarize, in this work we have addressed the magnetic properties of two-component superconductors in trigonal crystals with point group $D_{3d}$ symmetry. Starting from a general GL theory of trigonal two-component superconductors, we find that the upper critical field exhibits a \emph{sixfold} anisotropy in the basal plane, which is a discriminating property of two-component pairing. The sixfold anisotropy is a rare manifestation of discrete crystal symmetry, since effects of crystal anisotropy are typically obscured in GL theory by an emergent U(1) rotational symmetry. In addition, in this work we show that when a symmetry breaking field originating from, e.g., structural distortions selects a \emph{real} order parameter, $H_{c2}$ exhibits a twofold anisotropy with characteristic angular and temperature dependence.  

The recent NMR and specific heat measurements on Cu$_x$Bi$_2$Se$_3$, which reported spontaneously broken rotational symmetry, indicate that this material belongs to the class of superconductors with two-component pairing symmetry. Prominent other examples of materials with trigonal symmetry, which have attracted increasing attention recently, are the doped Bi$_2$Se$_3$ superconductors Sr$_x$Bi$_2$Se$_3$, Nb$_x$Bi$_2$Se$_3$, and Tl$_x$Bi$_2$Te$_3$. Our theory of in-plane anisotropy of upper critical field stands to contribute to uncovering the pairing symmetry of these superconductors, which remains to be determined. 

\section{Acknowledgments} 

We thank Anne de Visser, Shingo Yonezawa, Yoshi Maeno for discussions. This work is supported by the David and Lucile Packard Foundation (L.F.), the DOE Office of Basic Energy Sciences, Division of Materials Sciences and Engineering under Award No. DE- SC0010526 (V.K.), and the Netherlands Organization for Scientific Research (NWO) through a Rubicon grant (J.V.).


\pagebreak

\onecolumngrid

\setcounter{page}{1}

\begin{center}

{\bf \large Supplemental material for \\ ``Identification of nematic superconductivity from the upper critical field''}
\newline \newline
J\"orn W. F. Venderbos, Vladyslav Kozii, and Liang Fu \\
{\it Department of Physics, Massachusetts Institute of Technology,
Cambridge, MA 02139, USA}

\end{center}

\section{Landau theory of trigonal two-component superconductors\label{sec:gltheory}}

A Ginzburg-Landau (GL) theory of two-component superconductivity in a trigonal crystal with $D_{3d}$ point group symmetry is formulated in terms of the superconducting order parameters, which in turn are obtained from pairing potential. In the presence of spin-orbit coupling, when spin-rotation symmetry is broken, and the symmetry group is the symmetry group $D_{3d}$ of the crystal lattice, the superconducting pairing potential $\hat{\Delta}(\vec{k})$ is decomposed into irreducible representations of $D_{3d}$. In most cases, one is interested in a single pairing channel, i.e., a single representation of the symmetry group, which may be one- or multi-dimensional. The pairing matrix $\hat{\Delta}(\vec{k})$ corresponding to pairing channel $\Gamma$ takes the form~\cite{sigrist91_SM,volovik85_SM,mineevbook_SM}
\begin{gather}
\hat{\Delta}(\vec{k}) = \sum_{m} \eta_{\Gamma,m} \hat{\Delta}_{\Gamma,m} (\vec{k}),
\end{gather}
where $\Gamma$ labels the representation (i.e., pairing channel) and $m$ labels the components of the representation. The expansion coefficients $\eta_{\Gamma,m}$ are the superconducting order parameters and are complex scalars. In case of two-component superconductors the order parameter is the two-component complex number $\eta \equiv (\eta_1, \eta_2)^T$.

The GL free energy functional describing the two-component superconductor is then obtained as an expansion in the order parameters $\eta_{1,2}$ to given order,
\begin{gather}
F_{\Gamma}[\eta] =  \int d^3\vec{x} \; f_0 +  \int d^3\vec{x} \left(A(T)\sum_m |\eta_{\Gamma,m}|^2 + f^{(2n>2)}  \right), \label{eq:FE}
\end{gather}
which consists of terms fully invariant under the symmetries of the crystal. Therefore, the GL free energy depends on the crystal symmetry group as well as the pairing channel. A trigonal crystal with $D_{3d}$ point group symmetry admits two two-component representations, $E_g$ (even parity) and $E_u$ (odd parity), and in both cases the free energy density up to fourth order is given by
\begin{gather}
f = F/V = A( |\eta_1|^2 +|\eta_2|^2 )+B_{1} ( |\eta_1|^2 +|\eta_2|^2 )^2   + B_{2} | \eta_1^*\eta_2 - \eta_1\eta^*_2 |^2 , \label{eq:fetrig}
\end{gather}
in terms of the expansion coefficients $A\propto (T-T_c)$ and $B_{1,2}$. When the GL coefficient $B_2$ is positive ($B_2>0$) the order parameter $\eta$ is real and breaks rotational symmetry. In that case, a continuous degeneracy remains at fourth order, which is lifted by crystal anisotropy effects at sixth order. The sixth order contribution to the free energy density reads as
\begin{gather}
f^{(6)} = C_1[(\eta^*_+\eta_- )^3 + (\eta^*_-\eta_+ )^3] +C_{2} ( |\eta_1|^2 +|\eta_2|^2 )^3 + C_{3} ( |\eta_1|^2 +|\eta_2|^2 )| \eta_1^*\eta_2 - \eta_1\eta^*_2 |^2, \label{eq:fetrig6}
\end{gather}
(where $\eta_\pm  = \eta_1 \pm i \eta_2$) and it is the first term, proportional to $C_1$, which is responsible for lifting the continuous degeneracy and selecting three degenerate ground states related by threefold rotation.

We extend the GL theory to include terms representing spatial inhomogeneity: the gradient terms. The superconductor is a charged superfluid with charge $q=-2e$, and gradient terms are introduced by defining the covariant derivative as $D_i = -i\partial_i - q A_i$, where $A_i$ is the electromagnetic vector potential.

The gradient terms of the free energy expansion are obtained using the same representation theory recipe as before. The derivative components $(D_x,D_y)$ transform as $E_u$, and $D_z$ transforms as $A_{2u}$ of the $D_{3d}$ point group. These representations are used to form products with $(\eta_1,\eta_2)$, from which bilinears with the general structure
\begin{gather}
D_i X_{ia} \eta_a ,
\end{gather}
are obtained. Here $X_{ia}$ is a tensor transforming irreducibly under crystal symmetry (see Table~\ref{tab:irreps}). A gradient term is then simply given by $|D_i X_{ia} \eta_a|^2$, with an independent phenomenological stiffness constant for each distinct representation.


\begin{table}[t]
\centering
\begin{ruledtabular}
\begin{tabular}{ccc|ccc}
Symmetry  & Irreducible tensor $X_{ia}$ & $D_i X_{ia}\eta_a $  & $\times A_{1g}$  & $\times  A_{2g}$  &  $\times E_{g}$  \\ [4pt]
\hline
$A_{1g}$ &   $\tau^0_{ia} = \delta_{ia}$ &  $D_x \eta_1+D_y \eta_2 $  &  $A_{1g}$  & $A_{2g}$  &  $E_{g}$ \\ [3pt]
$A_{2g}$ &   $\tau^y_{ia}  $ & $-i(D_x \eta_2- D_y \eta_1 ) $   & $A_{2g}$  & $A_{1g}$  &  $E_{g}$ \\ [3pt]
$E_{g}$ &   $(\tau^x_{ia} , \tau^z_{ia})$ &  $(D_x \eta_2+ D_y \eta_1,D_x \eta_1- D_y \eta_2  )$  & $E_{g}$  & $E_{g}$  &  $A_{1g}+ A_{2g} + E_{g}$  \\ [3pt]
$E_{g}$ & $\delta_{iz}(\delta_{1a},\delta_{2a})$  &  $(D_z \eta_1,D_z \eta_2) $  & $E_{g}$  & $E_{g}$  &  $A_{1g}+ A_{2g} + E_{g}$ \\ [3pt]
\end{tabular}
\end{ruledtabular}
 \caption{Table listing the gradients terms present in a trigonal crystal with $D_{3d}$ point group symmetry. On the left the irreducible tensors $X_{ia}$ and corresponding bilinears are shown. On the right, a multiplication table of $D_{3d}$ representations is presented. This table assumes $(\eta_1,\eta_2)$ have $E_u$ symmetry, but the results for $E_g$ are obtained by exchanging the symmetry labels $g \leftrightarrow u$. }
\label{tab:irreps}
\end{table}

We now discuss all gradient terms of an odd-parity two-component superconductor in a trigonal crystal with $D_{3d}$ symmetry. We start from the gradient terms which have a continuous U(1) rotational symmetry with respect to the coordinates and order parameters individually, thus giving rise to an emergent U(1)$\times$U(1) symmetry. They are given by
\begin{gather}
f_{D,U(1)\times U(1)} = J_1 (D_i \eta_a)^*D_i \eta_a +J_2 \epsilon_{ij}\epsilon_{ab}(D_i \eta_a)^*D_j \eta_b   +J_3 (D_z \eta_a)^* D_z \eta_a ,  \label{eq:continuous}
\end{gather}
with gradient coefficients $J_{i}$ (and a sum over repeated $i,j$ and $a,b$ is understood, and $i,j=x,y$). As a result of the emergent U(1)$\times$U(1) symmetry, these gradient terms obscure the true discrete crystal rotation symmetry.

Whether or not the true crystal symmetry is reflected in the GL gradient expansion of multicomponent order parameters depends on the crystal system. For instance, in case of hexagonal symmetry (i.e., point group $D_{6h}$) the gradient terms are obtained by considering all irreducible tensors $X_{ia}$ of the hexagonal group. For the two-component representations $E_{1u,1g}$ the gradient terms are given by~\cite{sigrist91_SM}
\begin{gather}
f_D = K_1 |D_x \eta_1+D_y \eta_2 |^2 + K_2 |D_x \eta_2-D_y \eta_1 |^2 + K_3 (|D_x \eta_1-D_y \eta_2 |^2 +  |D_x \eta_2+D_y \eta_1 |^2) + K_4 (|D_z \eta_1 |^2 + |D_z \eta_2 |^2).  \label{eq:gradhexa}
\end{gather}
A very similar result is obtained for the $E_{2u,2g}$ representations. This can be rewritten to obtain (see also, e.g.,~\cite{sauls94_SM,zhitomirsky95_SM})
\begin{multline}
f_{D,U(1)\times U(1)} [J_1,J_2,J_3] +  f_{D,4}[J_4]  = J_1 (D_i \eta_a)^*D_i \eta_a +J_2 \epsilon_{ij}\epsilon_{ab}(D_i \eta_a)^*D_j \eta_b   +J_3 (D_z \eta_a)^* D_z \eta_a \\
 + J_4(\tau^z_{ij}\tau^z_{ab}+\tau^x_{ij}\tau^x_{ab})(D_i \eta_a)^*D_j \eta_b ,  \label{eq:gradhexa2}
\end{multline}
where summation over repeated indices is understood. The gradient constants $K_\alpha$ and $J_\alpha$ are related by
\begin{gather}
J_1 = \frac{K_1 + K_2}{2} + K_3, \quad J_2 = \frac{K_1 + K_2}{2} - K_3, \quad J_3 = K_4, \quad J_4 =  \frac{K_1 - K_2}{2}.
\end{gather}
The term proportional to $J_4$ has an emergent U(1) rotational symmetry: it is invariant under \emph{joint} rotations of coordinates and order parameters. (Note that the $J_4$ gradient term for  $E_{2u,2g}$ symmetry is given by $(\tau^z_{ij}\tau^z_{ab}-\tau^x_{ij}\tau^x_{ab})(D_i \eta_a)^*D_j \eta_b$, which also possesses an emergent U(1) symmetry.) Consequently, effects of crystal symmetry do not appear in the GL theory of hexagonal two-component superconductors.

When crystal symmetry is lowered to trigonal $D_{3d}$ symmetry, an additional gradient term arises which reflects the true $D_{3d}$ symmetry of the crystal. Table \ref{tab:irreps} lists all irreducible tensors $X_{ia}$ and their symmetries. With the help of this table it is straightforward to obtain all gradient terms. One observes that there are two distinct tensors with $E_g$ symmetry. As a result, a cross term of these two is an allowed gradient term. Specifically, the additional gradient term, defined as $f_{D,\text{trig}}[J_5]$ and referred to as trigonal anisotropy (gradient) term, takes the form~\cite{mineev_SM}
\begin{eqnarray}
f_{D,\text{trig}}[J_5]&=& J_5 \left[ (D_x \eta_2)^*D_z \eta_1 +(D_x \eta_1)^*D_z \eta_2 +  (D_y \eta_1)^*D_z \eta_1- (D_y \eta_2)^*D_z \eta_2  + c.c. \right] , \nonumber \\
&=& J_5 \left[ \tau^x_{ab}(D_x \eta_a)^*D_z \eta_b  +   \tau^z_{ab}(D_y \eta_a)^*D_z \eta_b + c.c. \right].    \label{eq:gradtrig}
\end{eqnarray}
The new gradient term is a consequence of trigonal anisotropy and is absent in hexagonal crystals, as is clear from Eq.~\eqref{eq:gradhexa2}. This is rooted in the fact that in trigonal crystals (i.e., only threefold rotations) angular momenta $L=0$ and $L=3$ are equivalent, whereas in hexagonal crystals with sixfold rotations these belong to distinct representations. 

In momentum space ($D_i \rightarrow q_i$) the trigonal gradient term can be expressed as $iq_z( q_-\eta^*_+\eta_- -q_+\eta^*_-\eta_+ )$, where $q_\pm = q_x \pm i q_y$ and similarly for $\eta_{1,2}$. The relative phases between $\eta_+$ ($q_+$) and $\eta_-$ ($q_-$) are determined by mirror symmetry: $\eta_1$ ($\eta_2$) is even (odd) under $x \rightarrow -x $. Even though the trigonal gradient term has $D_{3d}$ symmetry, and breaks the U(1) symmetry of Eq.~\eqref{eq:gradhexa2}, it possesses an emergent rotational symmetry: it is invariant under in-plane rotations of the order parameters and coordinates according to $q_+ \rightarrow q_+e^{2i\varphi}$, $\eta_+ \rightarrow \eta_+e^{-i\varphi}$. This, however, is not a physical symmetry.

For the gradient part of the free energy to be stable, the gradient coefficients $J_{1,\ldots,5}$ have to satisfy the stability conditions \cite{zhitomirsky90_SM,barash91_SM,zhitomirsky95_SM,mineev_SM}:
\begin{gather}
J_1+J_2 > 2|J_4|, \quad J_3 > 0, \quad J_3(J_1-J_2) > 2J^2_5.
\end{gather}

The GL theory of odd-parity two-component superconductors can be further extended by considering the coupling to other orders. In general, multi-component orders can be characterized by subsidiary order parameters. In case of the two-component trigonal superconductors, the subsidiary order parameters have symmetry $A_{2g}$ and $E_g$, and the corresponding bilinears of primary fields are given by
\begin{align}
A_{2g}  \;   \rightarrow \; \quad & \kappa = \eta^\dagger \tau^y \eta = - i (\eta_1^*\eta_2 - \eta_1\eta^*_2) =  \frac{1}{2}(| \eta_+|^2  - | \eta_-|^2 ), \\
 E_g \; \rightarrow \; \quad  & (N_1,N_2) = (\eta^\dagger \tau^z\eta, \eta^\dagger\tau^x \eta)= (| \eta_1|^2  - | \eta_2|^2, - \eta_1^*\eta_2 -\eta_1\eta^*_2) \label{eq:subs}
\end{align}
Here $\kappa$ and $(N_1,N_2)$ define the chiral and nematic subsidiary order parameters, respectively.

The coupling of the superconducting order to other orders is easily understood in terms of the subsidiary orders. A general magnetic order parameter $\vec{M}$ is a pseudo-vector and transforms as
\begin{gather}
M_z \rightarrow A_{2g} , \quad (M_x,M_y) \rightarrow E_{g},
\end{gather}
under trigonal symmetry $D_{3d}$. From Eq.~\eqref{eq:subs} we find that the only term that can couple to the magnetic order is of the form $| \eta_+|^2  - | \eta_-|^2$, giving rise to a coupling $ M_z ( | \eta_+|^2  - | \eta_-|^2) $. The magnetic order parameter $M_z$ can be thought of as a Zeeman field in the $z$ direction coupling to the electron (pseudo)spin as $\sim M_z\sigma_z$.

A structural deformation of the crystal, which leads to the breaking of rotational symmetry, can be expressed in terms of the strain tensor $u_{ij}$. Specifically, the uniaxial and shear strain components $(u_{xx} - u_{yy},2u_{xy} )$ transform as an $E_g$ doublet. As a result, these two strain tensor components can couple to the nematic superconductor associated with $E_g$ subsidiary order.

In terms of the coupling constants $g_M$ and $g_N$ we can write the contribution to the GL free energy coming from the magnetic and nematic coupling as
\begin{gather}
 g_M M_z ( |\eta_+|^2 - |\eta_-|^2 )  +  g_N [(u_{xx} - u_{yy})N_1 + 2  u_{xy} N_2 ]  \label{eq:mncouple}
\end{gather}
As is clear from the form of this term, the effect of symmetry breaking induced by the magnetic and nematic fields is to lift the degeneracy of the two-component pairing.

It is important to note that when we use the term ``strain'' here we very generally mean any nonmagnetic spin-rotation invariant order transforming as $E_g$.

\section{Solving Ginzburg-Landau equations for upper critical field $H_{c2}$}

In this section we present the solutions of the GL equations in the presence of a magnetic field $\vec{H}$, from which we find the expressions for the upper critical field $H_{c2}$. We first derive the GL equations from collecting the free energy contributions of Eqs.~\eqref{eq:FE}, \eqref{eq:gradhexa2}, and \eqref{eq:gradtrig}.

\subsection{Ginzburg-Landau equations}

We first derive the full set of GL equations governing the odd-parity two-component superconductor. To this end we write the total free energy as $F_{\text{tot}} = \int d^3\vec{x} \, f_{\text{tot}}  $, where $f_{\text{tot}} $ is the total free energy density. The free energy density has the following contributions
\begin{gather}
f_{\text{tot}} = f[A,B_1,B_2] +f_{D,U(1)\times U(1)} [J_1,J_2,J_3] + f_{D,4}[J_4]+ f_{D,\text{trig}}[J_5],
\end{gather}
where we make the functional dependence on GL coefficients explicit. The contribution from the electromagnetic field $\vec{A}$, given by $f_{EM} = (\vec\partial \times \vec A)^2/8\pi $ is neglected since it does not play a role in our theory. The GL equations follow from functional variation with respect to the fields and are given by
\begin{gather}
0= \frac{\delta F_{\text{tot}}}{\delta \eta^*_a}  = \frac{\delta F}{\delta \eta^*_a} + \frac{\delta F_{D,U(1)\times U(1)} }{\delta \eta^*_a} + \frac{\delta F_{D,4}}{\delta \eta^*_a}+\frac{\delta F_{D,\text{trig}}}{\delta \eta^*_a}.
\end{gather}
Starting with the homogeneous contribution to the free energy, this simply leads to
\begin{gather}
\frac{\delta F}{\delta \eta^*_a} = A\eta_a + 2 B_1\eta^\dagger  \eta \eta_a -2B_2 (\eta^*_1\eta_2- \eta^*_2\eta_1) \epsilon_{ab} \eta_b.
\end{gather}
For the gradient term $F^c_D$ one finds the contribution to the GL equations as
\begin{gather}
 \frac{\delta F_{D,U(1)\times U(1)}}{\delta \eta^*_a} = J_1 (D^2_x+D^2_y) \eta_a +\epsilon_{ab}J_2  (\vec{D}\times \vec{D})_z \eta_b+ J_3 D_z^2 \eta_a . \end{gather}
For the gradient term $f_{D,4}$ with U(1) symmetry one simply finds
\begin{gather}
 \frac{\delta F_{D,4}}{\delta \eta^*_a} = J_4(\tau^z_{ij}\tau^z_{ab}+\tau^x_{ij}\tau^x_{ab}) D_iD_j \eta_b .
\end{gather}
To conclude, the trigonal anisotropy term $f_{D,\text{trig}}$ gives a contribution to the GL equations which reads as
\begin{gather}
 \frac{\delta F_{D,\text{trig}}}{\delta \eta^*_a} = J_5\left[\tau^x_{ab}\{D_z,D_x \}+ \tau^z_{ab}\{D_z,D_y \} \right]\eta_b
\end{gather}

\subsection{The case $\vec{H} = H \hat{z}$\label{ssec:HinZ}}

We consider the case of a magnetic field applied in the $z$-direction (i.e., perpendicular to the basal plane of the crystal) given by $\vec{H} = H \hat{z}$. The corresponding vector potential in the symmetric gauge is $\vec{A} = -H \vec{x}\times \hat{z}/2 = - H \epsilon_{ij} x_j/2$, whereas in the Landau gauge $\vec{A}= H x \hat{y}$.

The GL equations (collecting the results $J_1$-$J_4$) for the order parameter $\eta$ read
\begin{gather}
-A\eta_a  = J_1 (D_x^2 + D_y^2 )\eta_a + J_3 D_z^2 \eta_a + J_2 \epsilon_{ab}[D_x,D_y]\eta_b + J_4\left[ (D^2_x-D^2_y)\tau^z_{ab}\eta_b + \{D_x,D_y\} \tau^x_{ab}\eta_b \right] \label{eq:GLHfield}.
\end{gather}
It will be beneficial to perform a basis transformation to positive and negative angular momentum combinations $\eta_\pm  = \eta_1 \pm i\eta_2$. The GL equations in this basis take the form
\begin{eqnarray}
-A\eta_+  & = & J_1 (D_x^2 + D_y^2 )\eta_+ + J_3 D_z^2 \eta_+ -i J_2 [D_x,D_y]\eta_+  + J_4 (D^2_x-D^2_y)\eta_- +i J_4 \{D_x,D_y\} \eta_- \nonumber \\
-A\eta_-  & = & J_1 (D_x^2 + D_y^2 )\eta_- + J_3 D_z^2 \eta_- + i J_2 [D_x,D_y]\eta_-  + J_4 (D^2_x-D^2_y)\eta_+ -i J_4 \{D_x,D_y\} \eta_+
\end{eqnarray}
In addition to an order parameter change of basis, we define the covariant derivatives $D_\pm = D_x \pm i D_y$. In terms of the angular momentum basis for the $D_i$ operators, the operators appearing in the GL equations are given by
\begin{gather}
D_x^2 + D_y^2  = \frac{1}{2} (D_+D_- + D_-D_+ ), \quad D_x^2 - D_y^2  = \frac{1}{2} (D_+D_+ + D_-D_- ), \nonumber \\
 [D_x,D_y] = \frac{i}{2} (D_+D_- - D_-D_+ ), \quad \{D_x,D_y\} = \frac{-i}{2} (D_+D_+ - D_-D_- )
\end{gather}
The problem of solving the GL equations is equivalent to the problem of two-component fermions in a magnetic field. To see this, we note that $[D_x,D_z]   =[D_y,D_z]   =0 $ and $[D_x,D_y]   =  iq \epsilon_{ij}\partial_iA_j = -2ie H = -i / l_b^2$. The latter commutation relation is the well-known commutation relation for momentum components in a magnetic field. We have defined a magnetic length $l_b = 1/\sqrt{2eH}$. Based on the commutation relation we define raising and lowering operators as $\Pi_\pm = l_b D_\pm /\sqrt{2}$ and substitute the expression in the GL equations to obtain
\begin{gather}
-A \begin{pmatrix} \eta_+ \\  \eta_- \end{pmatrix} = \frac{1}{l_b^2} \begin{pmatrix} J_1\{\Pi_-,\Pi_+ \} + J_2 [\Pi_+,\Pi_- ]& 0\\ 0& J_1\{\Pi_-,\Pi_+ \} -J_2[\Pi_+,\Pi_- ]   \end{pmatrix}\begin{pmatrix} \eta_+ \\  \eta_- \end{pmatrix}  +J_3 \begin{pmatrix}D^2_z&0 \\ 0&D^2_z \end{pmatrix} \begin{pmatrix} \eta_+ \\  \eta_- \end{pmatrix}.
\end{gather}
The $J_4$ term (and $J_5$ term) has been suppressed for the moment. The raising and lowering operators obey $ [\Pi_-,\Pi_+ ] = 1$ by definition. Furthermore, we see that inhomogeneity in the $z$-direction can only increase the (energy) eigenvalues and therefore corresponds to lower $H_{c2}$. We take $D_z\eta=0$. Then, including the $J_4$ term, the GL equations read
\begin{gather}
-A \begin{pmatrix} \eta_+ \\  \eta_- \end{pmatrix} = \frac{1}{l_b^2} \begin{pmatrix} J_1\{\Pi_-,\Pi_+ \} - J_2& 2J_4 \Pi^2_+ \\ 2J_4 \Pi^2_- & J_1\{\Pi_-,\Pi_+ \} +J_2 \end{pmatrix}\begin{pmatrix} \eta_+ \\  \eta_- \end{pmatrix}  .
\end{gather}
It may be simply checked that this system of equations is solved by either of the following solutions
\begin{gather}
 \begin{pmatrix} \eta_+ \\  \eta_- \end{pmatrix}  = \begin{pmatrix} \ket{n=0}  \\ 0   \end{pmatrix} , \qquad  \begin{pmatrix} \eta_+ \\  \eta_- \end{pmatrix}  = \begin{pmatrix} \alpha \ket{n+2}  \\  \beta \ket{n} \end{pmatrix},
\end{gather}
where $\ket{n}$ are harmonic oscillator eigenfunctions with the property $\Pi_+\Pi_- \ket{n} = n\ket{n}$. The corresponding expressions for the upper critical field are given by
\begin{gather}
H_{c2} = -A[2e(J_1 - J_2)]^{-1}, \qquad H_{c2} = -A\left[2e\left(J_1(2n+3) - \sqrt{(2J_1-J_2)^2 + 4J^2_4(n+1)(n+2)}  \right)\right]^{-1},
\end{gather}
and the actual physical upper critical field is given by the largest of the two values, with $n=0$ in the latter~\cite{zhitomirsky95_SM}.


As a next step, we consider the trigonal anisotropy term with gradient constant $J_5$. Rewriting it in terms of the $\eta_\pm$ basis and $\Pi_\pm$ operators it takes the form
\begin{gather}
 \frac{\sqrt{2}J_5}{l_b} \begin{pmatrix} 0 & i \{ D_z,\Pi_+ \}  \\ -i \{D_z,\Pi_- \}   & 0  \end{pmatrix}.
\end{gather}
Trigonal anisotropy in the GL equations thus couples the raising and lowering operators to $D_z$, and does not affect any homogeneous solution defined by the condition $D_z\eta = 0$. It is possible that an inhomogeneous solution has lower energy and therefore higher $H_{c2}$. To investigate this we take $D_z\eta = q_z \eta$. Setting $J_4=0$ as a first step, this leads to the system of eigenvalue equations
\begin{gather}
-A \begin{pmatrix} \eta_+ \\  \eta_- \end{pmatrix} = \frac{1}{l_b^2} \begin{pmatrix} J_1\{\Pi_-,\Pi_+ \} - J_2+ J_3\tilde{q}^2_z& i \sqrt{2}J_5\tilde{q}_z\Pi_-\\ -i \sqrt{2} J_5\tilde{q}_z \Pi_+ & J_1\{\Pi_-,\Pi_+ \} +J_2 + J_3\tilde{q}^2_z \end{pmatrix}\begin{pmatrix} \eta_+ \\  \eta_- \end{pmatrix}  .
\end{gather}
where we have defined the dimensionless momentum $\tilde{q}_z = l_bq_z$. It is easy to see that wave functions of the form
\begin{gather}
  \begin{pmatrix} \eta_+ \\  \eta_- \end{pmatrix}  = \begin{pmatrix} \alpha \ket{n}  \\  \beta \ket{n+1} \end{pmatrix},
\end{gather}
will solve this system. From the equations for $\alpha,\beta$ we find the eigenvalues and consequently $H_{c2}$ is given by
\begin{gather}
H_{c2} = |A|\left[2e\left(2J_1(n+1) +J_3\tilde{q}^2_z - \sqrt{(J_1+J_2)^2 + 2J^2_5\tilde{q}^2_z(n+1)}  \right)\right]^{-1}.
\end{gather}
The largest $H_{c2}$ corresponds to the $n=0$ solution. Such solution for $H_c2$ should be compared to the solution of the homogeneous case, i.e., choosing $\tilde{q}_z=0 $ to begin with. That solution is simply given by
\begin{gather}
H_{c2} = |A|[2e(J_1 - J_2)]^{-1}, 
\end{gather}
If this is larger, then a homogeneous solution gives the correct $H_{c2}$. For realistic values of the gradient coefficients, the homogeneous solution always leads to a larger $H_{c2}$.


\subsection{The case $\vec{A} = Hz (\sin \theta \hat{x}  -\cos\theta \hat{y})$\label{ssec:HinXY}}

Next, we come to the case of magnetic field in the basal plane: $\vec{H} = H(\cos\theta, \sin\theta,0)^T $. The corresponding vector potential configuration is given by $\vec{A} = Hz (\sin \theta, -\cos\theta,0)^T$, and in this gauge we have
\begin{gather}
D_x = -i\partial_x +2e Hz\sin\theta , \qquad D_y = -i\partial_y -2e Hz\cos\theta,
\end{gather}
in addition to $D_z = -i \partial_z$. Since the field is directed along a unit vector $(\cos\theta, \sin\theta,0)^T$ in the $xy$ plane, it is convenient to rotate the covariant derivative operators so that they are aligned with and orthogonal to the field direction. We define
\begin{gather}
 \begin{pmatrix} D_{\parallel} \\  D_{\perp} \end{pmatrix}  =  \begin{pmatrix} \cos\theta & \sin\theta \\ \sin\theta & -\cos\theta \end{pmatrix}   \begin{pmatrix} D_x \\  D_y \end{pmatrix} ,
\end{gather}
where $D_{\parallel}$ is along the field and $D_{\perp}$ is orthogonal to the magnetic field. As a result, $D_{\parallel}$ plays the same role as $D_z$ when the field is applied along the $z$-axis (see Sec.~\ref{ssec:HinZ}), and we have that $[D_{\parallel},  D_{\perp} ] = [D_{\parallel},  D_z ] =0$. The two components $ D_{\perp}$ and $D_z$ do not commute and define the magnetic algebra, i.e., $[ D_z, D_{\perp} ] =-2ie H$. In order to solve the GL Eq.~\eqref{eq:GLHfield} we reexpress them in terms of $D_{\parallel,\perp}$. The quadrature simply transforms as $D_x^2 + D_y^2  = D^2_{\parallel} + D^2_{\perp}$, and we find for the two ``$d$-wave'' components
\begin{gather}
 D_x^2 - D_y^2  = \cos 2\theta (D^2_{\parallel} - D^2_{\perp})  +  \sin 2\theta \{D_{\parallel},D_{\perp}\} , \nonumber \\
 \{D_x,D_y\} =   \sin 2\theta (D^2_{\parallel} - D^2_{\perp})  - \cos 2\theta   \{D_{\parallel},D_{\perp}\} .
\end{gather}
Note that $[D_x,  D_y ] =0$. Following the standard approach, we assume homogeneity in the direction along the field, i.e., $D_{\parallel}\eta = 0$, implying that we can ignore all terms containing $D_{\parallel}$.

The trigonal contribution to the GL equations takes the form
\begin{gather}
J_5\left[  \{D_z,D_x\} \tau^x_{ab}\eta_b + \{D_z,D_y\} \tau^z_{ab}\eta_b \right] ,
\end{gather}
which, after rewriting in terms of $D_{\parallel,\perp}$ and neglecting $D_{\parallel}$ becomes
\begin{gather}
J_5\left[  - \cos \theta  \tau^z_{ab}\eta_b + \sin \theta \tau^x_{ab}\eta_b \right]\{D_z,D_{\perp}\} . \label{eq:J5}
\end{gather}
We can now write the system of GL equations as a matrix equation for $\eta = (\eta_1,\eta_2)^T$ and find
\begin{gather}
-A  \eta = \left[ (J_1D^2_{\perp} + J_3 D^2_z)I - J_4D^2_{\perp} ( \cos 2\theta \tau^z+ \sin 2\theta \tau^x) +  J_5 \{D_z,D_{\perp}\} ( -\cos \theta \tau^z+ \sin \theta \tau^x)\right] \eta  \label{eq:trigonalsystem},
\end{gather}
where $I$ is the identity matrix. It is convenient to diagonalize the $J_4$ term by performing the following rotation
\begin{gather}
 \begin{pmatrix} \eta_1 \\  \eta_2 \end{pmatrix}  =  \begin{pmatrix} \cos\theta & - \sin\theta \\ \sin\theta & \cos\theta \end{pmatrix}   \begin{pmatrix} f_1 \\  f_2 \end{pmatrix}.
\end{gather}
In terms of the rotated order parameters $(f_1,f_2)^T$ the GL equations read
\begin{gather}
-A  \begin{pmatrix} f_1 \\  f_2 \end{pmatrix} = \begin{pmatrix} J_3D^2_z +(J_1-J_4)D^2_{\perp} & 0 \\ 0 &J_3D^2_z +(J_1+J_4)D^2_{\perp}\end{pmatrix}\begin{pmatrix} f_1 \\  f_2 \end{pmatrix}  +  J_5 \{D_z,D_{\perp}\} \begin{pmatrix} -\cos 3\theta &  \sin 3\theta \\ \sin 3\theta & \cos 3\theta \end{pmatrix} \begin{pmatrix} f_1 \\  f_2 \end{pmatrix}    \label{eq:trigonalrotated}.
\end{gather}
To solve this set of equations we employ the commutation relations of the operators $D_z$ and $D_{\perp}$, which give rise to a magnetic algebra equivalent to the case of $D_{x,y}$ when the field is applied in the $z$-direction (see Sec.~\ref{ssec:HinZ}). Specifically, we have $[ D_z, D_{\perp} ] =-i /l_b^2$, which implies that $D_z$ and $D_{\perp}$ satisfy a canonical commutation relation. It will be convenient to define effective ``momentum'' and ``position'' operators $\hat{P} = l_b D_z$ and $\hat{X} = l_b D_{\perp}$, so that $[\hat{X},\hat{P} ]= i $.

The appearance of the anti-commutator $\{D_z,D_{\perp}\} \sim \{ \hat{X},\hat{P} \} $ in Eq.~\eqref{eq:trigonalrotated} complicates this set of equations significantly, as it is not diagonal in the basis of harmonic oscillator states defined by the raising and lowering operators obtained from $\hat{X},\hat{P}$. Therefore, the presence of \emph{two} gradient terms originating from (trigonal) crystal anisotropy, the $J_4$ and $J_5$ terms, does not allow for a straightforward exact solution for arbitrary gradient coefficients.

To proceed, we map out the consequences of trigonal anisotropy terms on $H_{c2}$, starting from a number of limiting cases. In doing so, we obtain the full functional dependence of the upper critical field on the gradient coefficients. We will start by considering the cases $J_5=0$ (as applicable to hexagonal crystals), and $J_4=0$, i.e., when only the trigonal gradient coefficient $J_5$ is present. Then, we introduce the neglected terms perturbatively.

\begin{table}[t]
\centering
\begin{ruledtabular}
\begin{tabular}{ccl}
Symbol  & Definition & Physical meaning     \\ [4pt]
\hline
$l_b $ &   $1/l_b^2 = 2eH$ &  Magnetic length    \\ [4pt]
$\omega$ &   $\sqrt{J_1/J_3}$ &  Cyclotron frequency of the system  \\ [1pt]
& &  without crystal anisotropy effects   \\ [4pt]
$\omega_1$, $\omega_2$ &   $\omega\sqrt{1-|J_4|/J_1}$, $\omega\sqrt{1+|J_4|/J_1}$  & Cyclotron frequencies in presence of $J_4$   \\ [4pt]
$\omega_\pm$ &   $(\omega_2 \pm \omega_1)/2$ &  Sum and difference of frequencies    \\ [2pt]
& &  in presence of gradient term $J_4$   \\ [4pt]
$\hat{X}$, $\hat{P}$ & $l_b D_{\perp}$, $l_bD_z$  & Effective position and momentum operators    \\ [2pt]
&  & with canonical commutation relation     \\ [4pt]
$\Pi_\pm$ & $  (\hat{P} + i \omega\hat{X})/\sqrt{2\omega}$  &  Raising and lowering operators for Landau levels    \\ [2pt]
&  &  with cyclotron frequency $\omega$    \\ [4pt]
$\widetilde{\Pi}_\pm$ & $  e^{\pm i\pi/4}\Pi_\pm$  &  Rotated raising and lowering operators    \\ [4pt]
$\Xi_\pm$ & $  (\hat{P} + i \omega_1\hat{X})/\sqrt{2\omega_1}$  &  Raising and lowering operators for Landau levels   \\ [2pt]
&  &  with cyclotron frequency $\omega_1$    \\ [4pt]
\end{tabular}
\end{ruledtabular}
 \caption{This Table lists the definitions used in this section where we calculate the upper critical field in the basal plane. We note that in the Supplemental Material we define all cyclotron frequencies to be dimensionless by pulling out a factor $J_3$. We use a different convention in the Main Text.  }
\label{tab:defstrig}
\end{table}

\subsubsection{The hexagonal symmetry case ($J_5=0$)\label{ssec:hexagonal}}

This is the simplest case, as it reduces to the case of hexagonal symmetry (rather than trigonal symmetry, we are effectively assuming additional artificial symmetry) and the result is well-known~\cite{burlachkov85_SM}. It is nevertheless helpful to express the result in the present language. The system to solve takes the form
\begin{gather}
-A  \begin{pmatrix} f_1 \\  f_2 \end{pmatrix} =  \frac{J_3}{l^2_b} \begin{pmatrix} \hat{P}^2 + \omega^2_1 \hat{X}^2  & 0 \\ 0 &\hat{P}^2 + \omega^2_2 \hat{X}^2  \end{pmatrix}\begin{pmatrix} f_1 \\  f_2 \end{pmatrix} \label{eq:J4only} ,
\end{gather}
in terms of the position and momentum operators. The system is diagonal, with two different cyclotron frequencies given by
\begin{gather}
\omega^2_1 = \frac{J_1-J_4}{J_3}, \qquad \omega^2_2 = \frac{J_1+J_4}{J_3}.
\end{gather}
Hence, this system gives rise to two series of Landau levels. The upper critical field is simply determined by the lowest Landau level solution of the series with the smallest cyclotron frequency. Specifically, the upper critical field is given by
\begin{gather}
\widetilde H_{c2} = \max_{i=1,2} \{  \frac{-A}{2eJ_3\omega_i} \}. \label{eq: Hc2hexagonal}
\end{gather}

We note that here, in the Supplemental Material we define the cyclotron frequencies $\omega_1,$ $\omega_2$ (and similarly for $\omega,$ $\omega_{\pm}$ below) in a dimensionless fashion. A different convention is used in the Main Text, where an overall factor $J_3$ is absorbed in the cyclotron frequencies.

\subsubsection{The case when $J_4=0$ (only $J_5$)\label{ssec:trigonal}}

As a next case, we make the assumption that $J_4=0$, yet $J_5$ is nonzero, which may be called pure trigonal case. For this case, it is helpful to go back to equation~\eqref{eq:trigonalsystem}, which is expressed in terms of the $\eta$ variables. The system to solve is then given by
\begin{gather}
-A  \eta = \left[ (J_1D^2_{\perp} + J_3 D^2_z)I  +  J_5 \{D_z,D_{\perp}\} ( -\cos \theta \tau^z+ \sin \theta \tau^x)\right] \eta  ,
\end{gather}
and it is convenient to perform an order parameter rotation defined by
\begin{gather}
 \begin{pmatrix} \eta_1 \\  \eta_2 \end{pmatrix}  =  \begin{pmatrix} \cos(\theta/2) &  \sin(\theta/2) \\ - \sin(\theta/2) & \cos(\theta/2) \end{pmatrix}   \begin{pmatrix} f_1 \\  f_2 \end{pmatrix}.
\end{gather}
Performing this rotation we are left with a diagonal system which takes the form
\begin{gather}
-A  \begin{pmatrix} f_1 \\  f_2 \end{pmatrix} = \begin{pmatrix} J_1D^2_{\perp} + J_3 D^2_z -J_5 \{D_z,D_{\perp}\}& 0 \\ 0 &J_1D^2_{\perp} + J_3 D^2_z +J_5 \{D_z,D_{\perp}\} \end{pmatrix}\begin{pmatrix} f_1 \\  f_2 \end{pmatrix}    .
\end{gather}
The diagonal system left to solve is rather different from the previous case (i.e., $J_4 \neq 0, J_5=0$). This becomes apparent when we rewrite it in terms of $\hat{X}$ and $\hat{P}$,
\begin{gather}
-A  \begin{pmatrix} f_1 \\  f_2 \end{pmatrix} =   \frac{J_3}{l^2_b}  \begin{pmatrix} \omega^2 \hat{X}^2 +  \hat{P}^2 -J_5 \{\hat{X},\hat{P}\} /J_3& 0 \\ 0 & \omega^2 \hat{X}^2 + \hat{P}^2+J_5 \{\hat{X},\hat{P}\}/J_3\end{pmatrix}\begin{pmatrix} f_1 \\  f_2 \end{pmatrix}    , \label{eq:J5only}
\end{gather}
with $\omega^2 = J_1/J_3$. Since this system is diagonal, we can focus on the diagonal entries individually. To find the eigenvalues of such system, we introduce the raising and lowering operators $\Pi_\pm  =  (\hat{P} + i \omega\hat{X})/\sqrt{2\omega}$, with $\Pi_+ = \Pi^\dagger_-$. We then have for the right-hand side of Eq. (\ref{eq:J5only})
\begin{gather}
 \omega\left(\Pi_+\Pi_- + \Pi_-\Pi_+  \right) \pm i\frac{J_5}{J_3} \left(\Pi^2_+ - \Pi^2_- \right) . \label{eq: rhs of J5only}
\end{gather}
To proceed, we redefine the raising and lowering operators by performing the following ``rotation'': $\widetilde{\Pi}_\pm = e^{\pm i \pi/4}\Pi_\pm $. It can be checked that this corresponds to a rotation in the space of operator variables $\hat{X}$ and $\tilde{P}$. Note that such rotation preserves the operator algebra. We then find that (\ref{eq: rhs of J5only}) takes the form
\begin{gather}
 \omega\left(\widetilde{\Pi}_+\widetilde{\Pi}_- +\widetilde{\Pi}_-\widetilde{\Pi}_+  \right) \pm \frac{J_5}{J_3} \left(\widetilde{\Pi}^2_+ + \widetilde{\Pi}^2_- \right) . \label{eq:operatorrotate}
\end{gather}
We now use squeezing operators (see Sec.~\ref{ssec:squeezing}) to bring the operator equation into a form that is diagonal in the occupation number basis corresponding to the raising and lowering operators. Specifically, we use the unitary transformation
\begin{gather}
 \begin{pmatrix} f_1 \\  f_2 \end{pmatrix}  =  \begin{pmatrix} \hat{S}_+ &0 \\  0 & \hat{S}_-  \end{pmatrix}   \begin{pmatrix} g_1 \\  g_2 \end{pmatrix},
\end{gather}
where $\hat{S}_\pm$ are squeezing operators with $\theta=0$ for $\hat{S}_+$ and $\theta=\pi$ for $\hat{S}_-$. The squeezing parameter $r$ is defined by the relation
\begin{gather}
\tanh 2r= \frac{J_5}{\sqrt{J_1J_3}}
\end{gather}
The new cyclotron frequency is given by
\begin{gather}
\omega' = \sqrt{ \omega^2 - J^2_5/J^2_3} = \omega\sqrt{1 - J^2_5/J^2_3\omega^2} =  \omega \sqrt{ 1 - J^2_5/J_1J_3} \label{eq:omegap}
\end{gather}
This gives for the upper critical field
\begin{gather}
\widetilde H'_{c2} = \frac{-A}{2eJ_3\omega'} =\frac{-A}{2e \sqrt{ J_1J_3 - J^2_5} },   \label{eq:exactJ5only}
\end{gather}
Note that this implies a stability condition on the value of $J_5$ in relation to $J_{1,3}$, since the system does not make sense if $J_5$ exceeds both $J_{1,3}$.

\subsubsection{$J_5$ as a perturbation\label{ssec:J5perturb}}

Let us now reconsider Eqs.~\eqref{eq:trigonalrotated} and ~\eqref{eq:J4only}. We want to consider the $J_5$ term as a perturbation to Eq.~\eqref{eq:J4only}. It is convenient to use squeezing operators to make the cyclotron frequencies match in Eq.~\eqref{eq:J4only}. Let us assume that $\omega_2 > \omega_1$ (the calculation for the opposite case is equivalent), and take $\omega_2 / e^{2r}  = \omega_1$. We then have
\begin{gather}
\hat{S}^\dagger(r) \left( \hat{P}^2 + \omega^2_2 \hat{X}^2 \right)\hat{S}(r)  =  e^{2r} \left( \hat{P}^2 + \omega^2_1 \hat{X}^2 \right),
\end{gather}
where we have used a squeezing operator with $\theta=\pi $ (see Sec.~\ref{ssec:squeezing}). With this transformation, we bring the matrix operator equation into the form
\begin{gather}
-A  \begin{pmatrix} f_1 \\  \hat{S}^\dagger(r) f_2 \end{pmatrix} = \frac{J_3}{l^2_b} \begin{pmatrix}  \hat{P}^2 + \omega^2_1 \hat{X}^2 & 0 \\ 0 &  e^{2r} ( \hat{P}^2 + \omega^2_1 \hat{X}^2 ) \end{pmatrix}\begin{pmatrix} f_1 \\  \hat{S}^\dagger(r)  f_2 \end{pmatrix}  ,
\end{gather}
which does not yet include the $J_5$. The matrix operator in Eq. (\ref{eq:J5}) proportional to $J_5$, which we call $H'$ and consider as a perturbation, takes the following form after applying the squeezing transformation
\begin{gather}
H' = \frac{J_5/J_3}{l^2_b} \begin{pmatrix} -\cos 3\theta \{\hat{X},\hat{P}\} &  \sin 3\theta   \{\hat{X},\hat{P}\}  \hat{S}(r)  \\ \sin 3\theta   \hat{S}^\dagger(r) \{\hat{X},\hat{P}\}   & \cos 3\theta \{\hat{X},\hat{P}\}  \end{pmatrix}  \label{eq:J5perturbation}
\end{gather}

Let us first consider the diagonal matrix elements. We recognize that they have the same structure as in the case considered above ($J_4=0$, $J_5 \neq 0$). They can therefore be diagonalized in the same manner. We make the substitutions $\hat{P}^2 + \omega^2_1 \hat{X}^2 \rightarrow \omega_1 ( \widetilde{\Pi}_+\widetilde{\Pi}_- +\widetilde{\Pi}_-\widetilde{\Pi}_+  )$, $\{\hat{X},\hat{P}\}  \rightarrow -(\widetilde{\Pi}^2_+ + \widetilde{\Pi}^2_-)$ and $\hat{S}(r,\theta=\pi) \rightarrow \hat{S}(r,\theta=\pi/2)$, where in the latter case the squeezing operator is defined in terms of $\widetilde{\Pi}_\pm$.

We can immediately deduce an expression for $H_{c2}$ in case $r \gg 1$, i.e., $\omega_1 \ll \omega_2$. In that case we can ignore the coupling between the two series of Landau levels. All we are then left with is diagonalizing the matrix entry corresponding to $\omega_1$. This was achieved in the previous section and we simply substitute $J_5 \rightarrow J_5\cos 3\theta$. The cyclotron frequency $\omega_1$ transforms in the same way as $\omega$ (see above) and we obtain
\begin{gather}
\omega'_1 = \sqrt{ \omega_1^2 - J^2_5\cos^2 3\theta/J^2_3} = \sqrt{ (J_1-|J_4|)/J_3 - J^2_5\cos^2 3\theta/J^2_3}.
\end{gather}
As a result, in the limit $\omega_1 \ll \omega_2$ the upper critical field is given by
\begin{gather}
H_{c2} = \frac{-A}{2eJ_3\omega'_1} =\frac{-A}{2e \sqrt{ (J_1-|J_4|)J_3 - J^2_5\cos^2 3\theta} }.
\end{gather}
It is important to note that this limit corresponds to $(J_1 - |J_4|)/(J_1+|J_4|) \ll 1 $ with the hard constraint $ |J_5| < \sqrt{(J_1-|J_4|)J_3}$, which also should be considered a rather unphysical limiting case.

Let us now explore the case $\omega_1 \sim e^{2r} \omega_1 (= \omega_2)$. In this case we cannot ignore the coupling between the two series of Landau levels. It is convenient to work with raising and lowering operators $\Xi_\pm  =  (\hat{P} + i \omega_1\hat{X})/\sqrt{2\omega_1}$ (instead of the operators  $\Pi_\pm$, which are defined in terms of $\omega$). The eigenstates of the unperturbed system are given by [note that this is \emph{after} squeezing of the second component, i.e., in the basis $(g_1, g_2)^T \equiv (f_1,  \hat{S}^\dagger f_2)^T$ ]
\begin{gather}
\begin{pmatrix} \ket{n} \\ 0 \end{pmatrix}, \qquad \begin{pmatrix} 0 \\  \ket{n} \end{pmatrix},
\end{gather}
The ground state wave function of the unperturbed system is $(\ket{0}, 0)^T$. The operator part of the perturbation is given by $\{\hat{X},\hat{P}\} = -i (\Xi^2_+ - \Xi^2_- )$.

In general, when applying perturbation theory one must make two assumptions. First, one must obviously require that $J_5/J_3 \ll \omega_1 $ for the whole exercise to make sense. Second, one could naively think that whether to apply degenerate or non-degenerate perturbation theory depends on the smallness of $J_5/J_3$ as compared to the difference of the frequencies, i. e., whether the condition $J_5/J_3 \ll \omega_2 - \omega_1$ is satisfied or not. It turns out, however, that non-degenerate perturbation theory is sufficient even if the latter inequality is not satisfied. The reason for this is that the correction perturbative in $J_5$ remains small in the limit $\omega_1 \to \omega_2$. We confirm this conclusion explicitly applying degenerate perturbation theory.

In light of the latter discussion, we first consider non-degenerate perturbation theory. Given the operator part of the perturbation, we directly conclude that there is no first order correction to the cyclotron frequency. The second order correction to the cyclotron frequency is given by the sum of contributions from the two series of Landau level states, and it takes the form
\begin{eqnarray}
\delta \omega &=&  \frac{J^2_5\cos^2 3\theta}{J^2_3} \sum_{n \ge 1} \frac{| \bra{n} (\Xi^2_+ - \Xi^2_- ) \ket{0}|^2}{-2n\omega_1}+\frac{J^2_5\sin^2 3\theta}{J^2_3} \sum_{n \ge 0} \frac{| \bra{n} (\Xi^2_+ - \Xi^2_- )\hat{S} \ket{0}|^2}{\omega_1-(2n+1)\omega_2 }, \nonumber \\
&=&  -\frac{J^2_5\cos^2 3\theta}{2J^2_3\omega_1} +\frac{J^2_5\sin^2 3\theta}{J^2_3} \sum_{n \ge 0} \frac{| \bra{n} (\Xi^2_+ - \Xi^2_- )\hat{S} \ket{0}|^2}{\omega_1-(2n+1)\omega_2 }. \label{eq:PT1}
\end{eqnarray}
The presence of the squeezing operator complicates the summation in the last term. With the help of the properties of the squeezing operator, in particular the matrix elements with oscillator eigenstates (see Sec.~\ref{ssec:squeezing}), we find the matrix element of the numerator as
\begin{gather}
|\bra{n} (\Pi^2_+ - \Pi^2_- )\hat{S} \ket{0}|^2 = \left(\frac{n}{\sinh^2 r} -1 \right)^2 \tanh^2 r |\bra{n} \hat{S} \ket{0}|^2 =  \left(\frac{n}{\sinh^2 r} -1 \right)^2 \frac{n!}{(\tfrac{n}{2}!)^2 2^n}\frac{\tanh^{n+2} r}{\cosh r},
\end{gather}
Note that $n$ has to be even for a non-vanishing matrix element. The hyperbolic functions can be easily expressed in terms of the cyclotron frequencies $\omega_1$ and $\omega_2$. To simplify notation, and to express the results in terms of physically meaningful quantities, let us define $\omega_\pm = (\omega_2 \pm \omega_1)/2$. The hyperbolic functions then read
\begin{gather}
\sinh r  = \frac{\omega_-/\omega_+} {\sqrt{1- \omega^2_-/\omega^2_+}} , \quad \cosh r  = \frac{1} {\sqrt{1- \omega^2_-/\omega^2_+}} .
\end{gather}
Using these expressions we rewrite the sum in Eq.~\eqref{eq:PT1} as
\begin{gather}
\sum_{n \ge 0} \frac{| \bra{n} (\Xi^2_+ - \Xi^2_- )\hat{S} \ket{0}|^2}{\omega_1-(2n+1)\omega_2} = - \frac{ \omega^2_+}{ \omega^2_-}\left(1- \frac{ \omega^2_-}{ \omega^2_+} \right)^{5/2}\sum_{m \ge 0} \frac{(2m)!}{(m!)^2 4^m} \frac{(\omega_-/\omega_+)^{2m}}{4m\omega_2 +  2\omega_-}\left(2m- \frac{\omega^2_-/\omega^2_+}{1-\omega^2_-/\omega^2_+}  \right)^2,
\end{gather}
where we have changed variables from $n$ to $m$ since $n$ must be even. We notice that this sum is a function of the ratio of cyclotron frequencies $\omega_\pm$ only. Given this observation, we set $x \equiv \omega_-/\omega_+$ and we define a function $F_\pm(x,j)$, which depends on $x$ and integer $j$, as
\begin{gather}
F_\pm(x,j) =  \frac{ \left(1- x^2 \right)^{5/2}}{x^2}\sum_{m \ge j} \frac{(2m)!}{(m!)^2 4^m} \frac{x^{2m}}{2m\pm x(2m +1)}\left(2m- \frac{x^2}{1-x^2}  \right)^2. \label{eq:defF}
\end{gather}
The correction to the cyclotron frequency $\delta\omega$ of Eq.~\eqref{eq:PT1}, calculated using non-degenerate perturbation theory, can then be expressed as
\begin{gather}
\delta \omega =  -\frac{J^2_5}{2J^2_3}\left[\frac{\cos^2 3\theta}{\omega_1}+\frac{\sin^2 3\theta}{\omega_+} F_+(\tfrac{\omega_-}{\omega_+},0) \right]. \label{eq:resultnondegen}
\end{gather}
We point out that the function $F_{+}(x,0)$ is exactly the function $F(x)$ defined in the Main Text. It can be rewritten as
\be
F_+(x,0) = \frac{1-x}x \left[\sqrt{\frac{1+x}{1-x}} \, {}_2F_1\left( \frac12,  a; 1+  a; x^2 \right)  -1 \right],
\ee
where $a=x/2(1+x)$, and ${}_2F_1(\alpha, \beta; \delta ; \gamma )$ is a hypergeometric function. In our special case, it admits convenient integral representation:
\be
{}_2F_1\left( \frac12, a ; 1+a; x^2\right) = a \int_0^{\infty} \frac{dt \, e^{-at}}{\sqrt{1-x^2e^{-t}}}.
\ee

Including the correction (\ref{eq:resultnondegen}), the cyclotron frequency becomes $\omega_1  + \delta \omega $. The upper critical field $H_{c2}$ is always obtained from dividing $-A/2e$ by the cyclotron frequency, and therefore the upper critical field becomes, to lowest order in $J_5/J_3$,
\begin{gather}
H_{c2}(\theta) = \widetilde{H}_{c2}\left\{1 + \frac{J^2_5}{2J^2_3\omega_1}\left[\frac{\cos^2 3\theta}{\omega_1} + \frac{\sin^2 3\theta }{\omega_+} F_+(\omega_-/\omega_+,0) \right] \right\}, \label{eq:Hc2nondegen}
\end{gather}
where we have defined $\widetilde{H}_{c2} $ as the upper critical field in the absence of trigonal correction $J_5$, see Eq. (\ref{eq: Hc2hexagonal}).

Naively one would think that non-degenerate perturbation theory is no longer justified when $J_5/J_3 $ is larger than $\omega_2 - \omega_1$, in which case we must resort to (quasi)-degenerate perturbation theory. We will now present calculation using (quasi)-degenerate perturbation theory, and show that the result exactly equals the result of Eq.~\eqref{eq:resultnondegen}. We comment on this below.

In order to do (quasi)-degenerate perturbation theory we define the (quasi)-degenerate subspace by the states
\begin{gather}
\ket{1} = \begin{pmatrix} \ket{0} \\ 0 \end{pmatrix}, \qquad  \ket{2} = \begin{pmatrix} 0 \\  \ket{0} \end{pmatrix}. \label{eq:basisstates1}
\end{gather}
According to standard (quasi)-degenerate perturbation theory, the first and second order corrections to a Hamiltonian $H^{(0)}_{ij}$ are given by the general perturbative expression
\begin{gather}
H^{(0)}_{ij} + H^{(1)}_{ij} + H^{(2)}_{ij} = E_i\delta_{ij} + H'_{ij} + \frac{1}{2}\sum_k H'_{ik}H'_{kj}\left[ \frac{1}{E_i - E_k} +\frac{1}{E_j- E_k} \right], \label{eq:PTdegen1}
\end{gather}
where $E_i$ are the eigenvalues (i.e., cyclotron frequencies in the present problem) of the unperturbed problem, and $H'$ is the perturbation. In our current problem, the indices $i,j $ run over the basis states \eqref{eq:basisstates1}, and the perturbation $H'$ is given by Eq~\eqref{eq:J5perturbation}. Its matrix elements in the (quasi)-degenerate lowest Landau level subspace are given by
\begin{gather}
H^{(1)}_{ij} =   \frac{J_5\sin 3\theta}{J_3} \begin{pmatrix} 0 &  i\bra{0}\Xi^2_- \hat{S}\ket{0} \\ - i\bra{0}\hat{S}^\dagger \Xi^2_+ \ket{0}  & 0   \end{pmatrix} .
\end{gather}
To obtain the full correction to the cyclotron frequency to second order in $J_5/J_3$ we require the correction to the Hamiltonian up to order $J^2_5/J^2_3$, and therefore also need $H^{(2)}_{ij}$. Using the general expression of Eq.~\eqref{eq:PTdegen1} we find the diagonal matrix elements of $H^{(2)}_{ij}$ as
\begin{eqnarray}
H^{(2)}_{11} &=&   \frac{J^2_5}{J^2_3}\cos^2 3\theta \sum_{n \ge 1} \frac{| \bra{n} (\Xi^2_+ - \Xi^2_- ) \ket{0}|^2}{-2n\omega_1}+\frac{J^2_5}{J^2_3}\sin^2 3\theta \sum_{n \ge 1} \frac{| \bra{n} (\Xi^2_+ - \Xi^2_- )\hat{S} \ket{0}|^2}{\omega_1-(2n+1)\omega_2 }, \nonumber \\
&=&  -\frac{J^2_5}{2J^2_3\omega_1}\cos^2 3\theta +\frac{J^2_5}{J^2_3}\sin^2 3\theta \sum_{n \ge 1} \frac{| \bra{n} (\Xi^2_+ - \Xi^2_- )\hat{S} \ket{0}|^2}{\omega_1-(2n+1)\omega_2 }
\nonumber \\
&=&   -\frac{J^2_5}{2J^2_3}\left[\frac{\cos^2 3\theta}{\omega_1}+\frac{\sin^2 3\theta}{\omega_+} F_+(\tfrac{\omega_-}{\omega_+},1) \right],
\nonumber \\
H^{(2)}_{22} &=&   \frac{J^2_5}{J^2_3}\cos^2 3\theta \sum_{n \ge 1} \frac{| \bra{n} (\Xi^2_+ - \Xi^2_- ) \ket{0}|^2}{-2n\omega_2}+\frac{J^2_5}{J^2_3}\sin^2 3\theta \sum_{n \ge 1} \frac{| \bra{n} (\Xi^2_+ - \Xi^2_- )\hat{S} \ket{0}|^2}{\omega_2-(2n+1)\omega_1 }, \nonumber \\
&=&  -\frac{J^2_5}{2J^2_3\omega_2}\cos^2 3\theta +\frac{J^2_5}{J^2_3}\sin^2 3\theta \sum_{n \ge 1} \frac{| \bra{n} (\Xi^2_+ - \Xi^2_- )\hat{S} \ket{0}|^2}{\omega_2-(2n+1)\omega_1 }
\nonumber \\
&=&   -\frac{J^2_5}{2J^2_3}\left[\frac{\cos^2 3\theta}{\omega_2}+\frac{\sin^2 3\theta}{\omega_+} F_-(\tfrac{\omega_-}{\omega_+},1) \right],
\end{eqnarray}
In principle we also need the off-diagonal matrix elements, however, these will only contribute to cyclotron frequency correction at order $\sim J^4_5/J^4_3$ and we therefore ignore them. The Hamiltonian matrix to order $J^2_5/J^2_3$ can be cast into the general form $a I_2 + b\tau^z + c\tau^x + d \tau^y$ where $\tau^i$ are Pauli matrices and $a,b,c,d$ are real expansion coefficients. Then, the new cyclotron frequency $\omega'$ entering the expression for $H_{c2}$ is given by $\omega'  = a - \sqrt{b^2 + c^2 + d^2 }$. We find that the coefficients $a$ and $b$ are given by
\begin{eqnarray}
a &= & \omega_+ - \frac{J^2_5}{2J^2_3\omega_+} \left[\frac{\cos^2 3\theta}{1-(\omega_-/\omega_+)^2} + \sin^2 3\theta \frac{F_+ + F_-}{2} \right], \nonumber \\
b &=& -\omega_- - \frac{J^2_5}{2J^2_3\omega_+} \left[\cos^2 3\theta \frac{\omega_-/\omega_+}{1-(\omega_-/\omega_+)^2} + \sin^2 3\theta \frac{F_+ - F_-}{2} \right],
\end{eqnarray}
where $F_\pm$ are abbreviations for $F_\pm(\omega_-/\omega_+,1)$. The coefficients $c$ and $d$ are defined through the equation
\begin{gather}
c\tau^x + d \tau^y = \frac{J_5\sin 3\theta}{J_3} \begin{pmatrix} 0 &  i\bra{0}\Xi^2_- \hat{S}\ket{0} \\ - i\bra{0}\hat{S}^\dagger \Xi^2_+ \ket{0}  & 0   \end{pmatrix}.
\end{gather}
These matrix elements are further simplified using $\bra{0}\Xi^2_- \hat{S}\ket{0}  = -\tanh r \bra{0} \hat{S}\ket{0}$. Putting everything together, we evaluate the square root $\sqrt{b^2 + c^2 + d^2 } $ to second order in $J_5/J_3$ and find
\begin{gather}
\sqrt{b^2 + c^2 + d^2 }  = \omega_-  +  \omega_- \frac{J^2_5}{2J^2_3\omega^2_+}\left[\cos^2 3\theta \frac{1}{1-(\omega_-/\omega_+)^2}   + \sin^2 3\theta \frac{\omega_+}{\omega_-}\frac{F_+ - F_-}{2} +\sqrt{1-\frac{\omega^2_-}{\omega^2_+}} \sin^2 3\theta \right]  + \mathcal{O}(J^4_5/J^4_3)
\end{gather}
We then find the cyclotron frequency $\omega'$ to second order in $J_5/J_3$ as
\begin{gather}
\omega'=  (\omega_+ -  \omega_- )\left\{1 - \frac{J^2_5}{2J^2_3\omega^2_+}\left[\frac{\cos^2 3\theta}{(1-\omega_-/\omega_+)^2} + \frac{\sin^2 3\theta }{1-\omega_-/\omega_+}\left( F_+(\tfrac{\omega_-}{\omega_+},1) + \frac{\omega_-}{\omega_+}\sqrt{1-\frac{\omega^2_-}{\omega^2_+}}\right) \right] \right\}.
\end{gather}
For convenience we have separated $\omega_+ -  \omega_-  = \omega_1$, which is the cyclotron frequency of the unperturbed case with $J_5=0$. Next, we notice that the term $\omega_-(1-\omega^2_-/\omega^2_+)^{1/2}/\omega_+$ is precisely equal to the $m=0$ term in the definition of $F_+(\omega_-/\omega_+,0)$, and we can therefore add it to $F_+(\omega_-/\omega_+,1)$ to obtain $F_+(\omega_-/\omega_+,0)$. As a result, we arrive at the final expression for $H_{c2}$ containing corrections to order $J^2_5/J^2_3$
\begin{gather}
H_{c2} = \widetilde{H}_{c2}\left\{1 + \frac{J^2_5}{2J^2_3\omega^2_+}\left[\frac{\cos^2 3\theta}{(1-\omega_-/\omega_+)^2} + \frac{\sin^2 3\theta }{1-\omega_-/\omega_+} F_+(\tfrac{\omega_-}{\omega_+},0) \right] \right\}.
\end{gather}
Remarkably, this expression is precisely equal to the upper critical field of Eq.~\eqref{eq:Hc2nondegen}, which was obtained through non-degenerate perturbation theory.

\subsubsection{$J_4$ as a perturbation\label{ssec:J4perturb}}

In this limiting case, we start by reconsidering Eq.~\eqref{eq:J5only}. In order to treat the gradient term with coefficient $J_4$, we first perform the same rotation of order parameter variables. This yields
\begin{gather}
\frac{J_4}{J_3}\hat{X}^2 \begin{pmatrix} \cos 3\theta &  \sin 3\theta    \\ \sin 3\theta    & -\cos 3\theta \end{pmatrix}  \begin{pmatrix} f_1 \\  f_2 \end{pmatrix}
\end{gather}
In order to express this in the same operators that were used to solve Eq.~\eqref{eq:J5only}, we first have to rotate the raising and lowering operators [i.e., $\Pi_\pm \rightarrow \widetilde{\Pi}_\pm $, see Eq.~\eqref{eq:operatorrotate}] and then apply the squeezing transformation. After rotation we have $ \hat{X} \rightarrow (\hat{X} + \hat{P}/\omega)/\sqrt{2}$ and therefore $\hat{X}^2 \rightarrow (\hat{X}^2 + \hat{P}^2/\omega^2 + \{ \hat{X},\hat{P} \}/\omega )/2$. Applying the squeezing transformation leads to
\begin{gather}
 \begin{pmatrix} \hat{S}_- &0 \\  0 & \hat{S}_+  \end{pmatrix}  \frac{J_4}{2J_3}(\hat{X}^2 + \frac{1}{\omega^2}\hat{P}^2+ \frac{1}{\omega}\{ \hat{X},\hat{P} \}) \begin{pmatrix} \cos 3\theta &  \sin 3\theta    \\ \sin 3\theta    & -\cos 3\theta \end{pmatrix}     \begin{pmatrix} \hat{S}_+ &0 \\  0 & \hat{S}_-  \end{pmatrix} .
\end{gather}
We may then use the following properties (bearing in mind that $\hat{S}_-  =\hat{S}^\dagger_+$): $\hat{S}_- \hat{P} \hat{S}_+  = e^{-r}\hat{P}$, $\hat{S}_- \hat{X} \hat{S}_+  = e^{r}\hat{X}$. We then obtain
\begin{multline}
 \frac{J_4\cos 3\theta }{2J_3} \begin{pmatrix}(e^{2r}\hat{X}^2 + \frac{e^{-2r}}{\omega^2}\hat{P}^2+ \frac{1}{\omega}\{ \hat{X},\hat{P} \})  &  0    \\0    & - (e^{-2r}\hat{X}^2 + \frac{e^{2r}}{\omega^2}\hat{P}^2+ \frac{1}{\omega}\{ \hat{X},\hat{P} \}) \end{pmatrix}  + \nonumber \\
 \frac{J_4\sin 3\theta }{2J_3}\begin{pmatrix} 0  & (e^{2r}\hat{X}^2 + \frac{e^{-2r}}{\omega^2}\hat{P}^2+ \frac{1}{\omega}\{ \hat{X},\hat{P} \})\hat{S}^2_-   \\ \hat{S}^2_+ (e^{2r}\hat{X}^2 + \frac{e^{-2r}}{\omega^2}\hat{P}^2+ \frac{1}{\omega}\{ \hat{X},\hat{P} \})   & 0 \end{pmatrix}      .
\end{multline}
In order to do degenerate perturbation theory we introduce the harmonic oscillator basis states
\begin{gather}
\ket{1} = \begin{pmatrix} \ket{0} \\ 0 \end{pmatrix}, \qquad  \ket{2} = \begin{pmatrix} 0 \\  \ket{0} \end{pmatrix},
\end{gather}
and calculate the first order correction to the Hamiltonian given by Eq.~\eqref{eq:PTdegen1}. We use the following standard results: $\bra{0} \hat{X}^2 \ket{0} = 1/2\omega = \bra{0} \hat{P}^2 \ket{0} /\omega^2$, and $\bra{0} \{\hat{X}, \hat{P}\} \ket{0}=0 $. We find for the diagonal part of $H^{(1)}$
\begin{gather}
 \frac{J_4\cos 3\theta \cosh 2r}{2\omega J_3} \begin{pmatrix}1   &  0    \\0    & - 1 \end{pmatrix}       .
\end{gather}
The next step is to evaluate the expectation value of the off-diagonal elements. This is more involved as a result of the appearance of the squeezing operators. We find that the off-diagonal elements take the form
\begin{gather}
\bra{0}(e^{2r}\hat{X}^2 + \frac{e^{-2r}}{\omega^2}\hat{P}^2+ \frac{1}{\omega}\{ \hat{X},\hat{P} \})\hat{S}^2_- \ket{0}  =  \frac{1}{\omega}  \left( \frac{1+i \sinh 2r}{\cosh 2r} \right)\bra{0}\hat{S}^2_- \ket{0} .
\end{gather}
Using this, we find the correction to the cyclotron frequency $\omega'$, which was defined in Eq.~\eqref{eq:omegap}, as
\begin{gather}
\delta \omega' =  - \frac{J_4}{2 J_3} \sqrt{\cos^2 3\theta \frac{\cosh^2 2r}{\omega^2} +  \sin^2 3\theta \frac{|\bra{0}\hat{S}^2_- \ket{0}|^2 }{\omega^2}}  .
\end{gather}
This expression can be simplified by noting that $\hat{S}^2_-  = [\hat{S}(r,\theta=\pi)]^2 = \hat{S}(2r,\theta=\pi)$, and using the definition of $r$. One finds $\cosh 2r = \omega / \omega'$. In addition, the expectation value of the squeezing operator in the oscillator ground vacuum state is given by $|\bra{0}\hat{S}^2_- \ket{0}|^2 = 1/\cosh 2r$. Putting this all together we find the correction $\delta \omega' $ as
\begin{gather}
\delta \omega' =  - \frac{J_4}{2 J_3 \omega'} \sqrt{\frac{\omega'^3 }{\omega^3} + \cos^2 3\theta \left(1- \frac{\omega'^3 }{\omega^3} \right) }   .
\end{gather}
Consequently, the upper critical field to the first order in $J_4$ is given by
\begin{multline}
H_{c2}(\theta) = \widetilde{H}'_{c2} \left[1+\frac{J_4}{2 J_3 \omega'^2} \sqrt{\frac{\omega'^3 }{\omega^3} + \cos^2 3\theta \left(1- \frac{\omega'^3 }{\omega^3} \right) }  \right] \\ = - \frac{A}{2e\sqrt{J_1 J_3 -J_5^2}}\left[ 1+ \frac{|J_4| J_3}{2(J_1J_3-J_5^2)}\sqrt{\cos^2 3\theta + \sin^2 3\theta \left(1 - \frac{J_5^2}{J_1 J_3}  \right)^{3/2}} \right],
\end{multline}
where $\widetilde{H}'_{c2}$ is defined as the upper critical field in the absence of $J_4$ term, see Eq. (\ref{eq:exactJ5only}).


\subsubsection{Summary of the result for $H_{c2}$ in the presence of trigonal anisotropy}

For convenience and clarity of presentation, here we collect the main results of basal plane $H_{c2}$ calculations. In particular, we list the expressions for $H_{c2}$ in all limiting cases we considered, and state their ranges of applicability.

Taking $J_5=0$, the case which is applicable to a hexagonal crystal, the exact solution is 
\be
J_5 = 0 , \qquad H_{c2} = \frac{-A}{2e\sqrt{J_3(J_1-|J_4|)}}.
\ee
I$H_{c2}$ does not exhibit any angular dependence due to the emergent $U(1)$ symmetry of the GL equations.

Taking $J_4=0$, where $J_5$ is the only gradient coefficients in addition to $J_{1,2,3}$ the exact solution of the GL equations is
\be
J_4 = 0, \qquad H_{c2} = \frac{-A}{2e\sqrt{J_1J_3 - J_5^2}}. \label{Eq: exactJ5only}
\ee
Due to an emergent (but unphysical) rotational symmetry of the GL equations, $H_{c2}$ is isotropic (see Main Text).

We obtain the following approximate solution in the case where $|J_4|\approx J_1$ and $J_5$ satisfies the stability constraint $J_5 < \sqrt{J_3(J_1-|J_4|)}$
\be
J_1 - |J_4| \ll J_1 + |J_4|, \qquad H_{c2} = \frac{-A}{2e \sqrt{J_3(J_1-|J_4|) - J_5^2 \cos^2 3\theta}} . \label{Eq: largeJ4}
\ee

The most physical case arises when $J_4$ takes arbitrary values (but $|J_4|<J_1$ due to stability constraints) and $J_5$ is small. The expression for $H_{c2}$ reads as
\be
|J_5| \ll \sqrt{J_3 (J_1 - |J_4|)}, \qquad H_{c2}(\theta) = \frac{-A}{2e\sqrt{J_3(J_1-|J_4|)}} \left\{1 + \frac{J^2_5}{2J^2_3\omega_1}\left[\frac{\cos^2 3\theta}{\omega_1} + \frac{\sin^2 3\theta }{\omega_+} F(\omega_-/\omega_+) \right] \right\}, \label{Eq: J5pert}
\ee
where $F(x)$ is the function defined in Eq.~\eqref{eq:defF}, and definitions of $\omega,$ $\omega_{\pm}$, $\omega_{1,2}$ are listed in Table \ref{tab:defstrig}.

We also considered the opposite case, where $J_5$ is arbitrary and satisfies $J_5<\sqrt{J_1 J_3}$, and $J_4$ is small. The result is
\be
|J_4|J_3\ll J_1 J_3 - J_5^2 , \qquad H_{c2}(\theta) =  - \frac{A}{2e\sqrt{J_1 J_3 -J_5^2}}\left[ 1+ \frac{|J_4| J_3}{2(J_1J_3-J_5^2)}\sqrt{\cos^2 3\theta + \sin^2 3\theta \left(1 - \frac{J_5^2}{J_1 J_3}  \right)^{3/2}} \right]. \label{Eq: J4pert}
\ee
This reproduces the result of Ref.~\onlinecite{mineev_SM}. The advantage of this solution is that it can be obtained by first-order perturbation theory in $J_4$.

We can expand Eq.~\eqref{Eq: J5pert} in small $J_4$ and Eq.~\eqref{Eq: J4pert} in small $J_5$. In both cases the result is 
\be
|J_4|\ll J_1, \quad J_5^2\ll J_1 J_3, \qquad H_{c2} = -\frac{A}{2e\sqrt{J_1 J_3}}\left[ 1+ \frac{|J_4|}{2J_1} + \frac{J_5^2}{2J_1 J_3} + \frac{3 |J_4| J_5^2}{8J_1^2 J_3}\left(1+ \cos^2 3\theta \right)  \right].
\ee

Finally, we mention that in the case of large $|J_4|$, $J_1 - |J_4| \ll J_1 + |J_4|$ (corresponding to $\omega_1 \ll \omega_2$), Eq.~(\ref{Eq: J5pert}), which is correct for arbitrary $|J_4|<J_1$, reproduces to the first-order (in $J_5^2$) correction of Eq.~(\ref{Eq: largeJ4}):
\be
\frac{J_1 - |J_4|}{J_1+|J_4|} \ll 1, \quad J_5^2 \ll J_3 (J_1 - |J_4|), \qquad H_{c2} = \frac{-A}{2e\sqrt{J_3(J_1-|J_4|)}} \left( 1 + \frac{J_5^2 \cos^2 3\theta}{2J_3 (J_1- |J_4|)} \right).
\ee

\subsection{Upper critical field in the presence of symmetry breaking field}

We now come to the problem of calculating $H_{c2}$ in the presence of a symmetry breaking field which pins the order parameter along a preferred axis. As in the main text, we take the symmetry breaking field to have $\epsilon_{xx}-\epsilon_{yy}$ symmetry, which implies a contribution to the free energy given by $\delta\tau_{ij}^z\eta^*_i\eta_j$. We then solve the GL equations in the presence of an in-plane magnetic field (see Sec.~\ref{ssec:HinXY}) for two cases: (i) when only $J_4$ is present ($J_5=0$), and (ii) when both $J_4$ and $J_5$ are present.

\subsubsection{Symmetry breaking field $\delta$ and $J_4$ }

We first consider case of $J_4$ only, where trigonal anisotropy is set to zero~\cite{agterberg94_SM,agterberg95_SM}. Starting from Eq.~\eqref{eq:trigonalsystem} and adding the symmetry breaking field, the GL equations take the form
\begin{gather}
-A  \eta = \left[ (J_1D^2_{\perp} + J_3 D^2_z)I - J_4D^2_{\perp} ( \cos 2\theta \tau^z+ \sin 2\theta \tau^x) +  \delta\tau^z  \right] \eta  \label{eq:symbreak}.
\end{gather}
We follow the same approach as in prior parts of this section and rewrite the matrix equation in terms of position and momentum operators $\hat{X}$ and $\hat{P}$ to obtain
\begin{gather}
-\frac{A}{J_3}  \eta = \left[ \frac{1}{l^2_b}(\hat{P}^2 + \omega^2 \hat{X}^2 )I -  \frac{J_4}{J_3 l^2_b}\hat{X}^2 ( \cos 2\theta \tau^z+ \sin 2\theta \tau^x) +  \frac{\delta}{J_3} \tau^z  \right] \eta  \label{eq:symbreak2}.
\end{gather}
This system of equations is similar to what has been considered in the context of UPt$_3$~\cite{agterberg94_SM,agterberg95_SM}. In order to find the upper critical field, we simply project it into the lowest Landau level solutions $(\ket{0},0)^T$ and $(0,\ket{0})^T$.  This is justified as long as $J_4\ll J_1.$ Noting that $\bra{0}\hat{X}^2\ket{0} = 1/2\omega$, we find the following implicit equation for the upper critical field
\begin{gather}
-\frac{A}{J_3\omega}  =  \frac{1}{l^2_b}  - \sqrt{ \frac{J^2_4}{4\omega^4 J^2_3 l^4_b} \sin^2 2\theta + \left( \frac{\delta}{J_3\omega} - \frac{J_4}{2\omega^2 J_3 l^2_b}\cos 2\theta \right)^2  }.  \label{eq:symbreak3}
\end{gather}
To extract physical information from this equation we define the following quantities
\begin{gather}
x = \frac{J_4}{\omega^2 J_3 } = \frac{J_4}{J_1 }, \quad y = \frac{\delta}{J_3\omega} = \frac{\delta}{\sqrt{J_1J_3}}. \label{eq:defxy}
\end{gather}
Here $x$ is a measure of the strength of the hexagonal anisotropy $J_4$ as compared to the (unperturbed) cyclotron frequency, and $y$ is a measure the symmetry breaking field as compared to the cyclotron frequency. In terms of these parameters the equation reads
\begin{gather}
-\frac{A}{J_3\omega}    =  \frac{1}{l^2_b}  - \sqrt{ \frac{x^2}{4 l^4_b} \sin^2 2\theta + \left(y - \frac{x}{2 l^2_b}\cos 2\theta \right)^2  }.
\end{gather}
It is important to note that the magnetic length, and therefore the magnetic field, enters this expression in a non-trivial way. The physical implication is that the upper critical field will exhibit different behavior close to $T_c$ as compared to far below $T_c$. This may be understood from the fact that the symmetry breaking field splits the transitions of two order parameter components at zero field, as $T_c$ depends linearly on $\delta$ within the framework of GL theory.

We further rewrite the implicit equation for $H_{c2}$ to obtain
\begin{gather}
-\frac{A}{J_3\omega} =  \frac{1}{l^2_b}  - \sqrt{ \frac{x^2}{4 l^4_b} + y^2  - \frac{xy}{l^2_b}\cos 2\theta   }, \label{eq:quadraticHc2}
\end{gather}
or, in terms of the original variables,
\be
-\frac{A}{\sqrt{J_1 J_3}} = 2eH - \sqrt{\frac{e^2 H^2 J_4^2}{J_1^2} + \frac{\delta^2}{J_1 J_3} - \frac{ 2 e H J_4 \delta}{\sqrt{J_1^3 J_3}}\cos 2\theta}.
\ee
In the limit of small fields, expressed as  $x \ll y l^2_b $  (corresponding to $e H J_4\sqrt{J_3} \ll \delta \sqrt{J_1}$ ), we can ignore the first term in the square root and expand it in $x / yl^2_b$. In this case, we obtain the equation
\begin{gather}
-\frac{A}{J_3\omega} =  \frac{1}{l^2_b}  - | y |\left(1 - \frac{x}{2 y l^2_b}\cos 2\theta  \right), \label{eq:Hc2smallH}
\end{gather}
from which we obtained the upper critical field in the limit of small fields as
\begin{gather}
H_{c2}=  -\frac{A-|\delta| }{2eJ_3\omega}\left(1 - \frac{J_4\, \text{sgn}(\delta)}{2J_1}\cos 2\theta  \right).
\end{gather}

Since Eq.~\eqref{eq:quadraticHc2} is a quadratic equation for $1/l^2_b\sim H$, we can solve it for $H_{c2}$. To this end, we first define $-A /|\delta| = t = t(T)$, which is a function of temperature. We then find for $H_{c2}$

\begin{eqnarray}
2eH_{c2} &=&  \frac{x\,y\cos 2\theta - 2t|y|  - \sqrt{(x\,y\cos 2\theta - 2t|y|)^2-y^2 (x^2-4)(1-t^2)}}{(x^2/2-2)}  \\ 
&=& \frac{-(2AJ_1 + \delta J_4 \cos 2\theta) + \sqrt{(2J_1 A + \delta J_4 \cos 2\theta)^2 + (4J_1^2 - J_4^2)(A^2 - \delta^2)}}{(4J_1^2 - J_4^2)\sqrt{J_3/4J_1}}.
\end{eqnarray}
At this point, it is worth commenting on the variable $t$ [the parameters $x$ and $y$ were defined in Eq.~\eqref{eq:defxy}]. To make its significance explicit, we write explicitly $A = A'(T-T_c),$. In weak-coupling $A'$ is equal to (up to the constant of the order one) the density of states. Then, the transition temperature in the presence of the pinning field $\delta$  is shifted to $T_c^* = T_c + |\delta|/A',$ or $\Delta T_c = T_c^* - T_c = |\delta|/A'.$ In this notation, $t = -1 - (T-T_c^*)/\Delta T_c $, and $t$ takes the role of an effective temperature through $T=  T^*_c -(t+1)\Delta T_c$. Note that in the Main Text we have left the proportionality constant $A'$ implicit in the definition of $t$. Also note that in the superconducting state $t $ takes values $t \ge -1$. We thus observe that the upper critical field, in particular its angular dependence, is a function of temperature through $t = t(T) $.

Let us see how this is reflected in the $H_{c2}$ anisotropy coefficient $H_{c2}(\tfrac{\pi}{2})/H_{c2}(0)$. We find that it takes the form
\begin{gather}
\frac{H_{c2}(\tfrac{\pi}{2})}{H_{c2}(0)} =  \frac{-x\,\text{sgn}(y) -2t - |2+x\,\text{sgn}(y)t|}{x\,\text{sgn}(y) - 2t - |2-x\,\text{sgn}(y)t|}.
\end{gather}
Close to $T^*_c$, at $t<2/x$ (corresponds to $T>T_c-2J_1|\delta|/J_4 A'$), it can be shown to reduce to
\begin{gather}
\frac{H_{c2}(\tfrac{\pi}{2})}{H_{c2}(0)} =  \frac{1+x\, \text{sgn}(y) /2 }{1-x \,\text{sgn}(y) /2 } = \frac{1+J_4\, \text{sgn}(\delta) /2J_1 }{1-J_4\, \text{sgn}(\delta) /2J_1 },
\end{gather}
which is independent of temperature and in agreement with Eq.~\eqref{eq:Hc2smallH}. In contrast, when $t  >2/x$, i.e., at temperatures far below $T^*_c$, we find that [in case $\text{sgn}(y) >0 $]
\begin{gather}
\frac{H_{c2}(\tfrac{\pi}{2})}{H_{c2}(0)} =  -\frac{1+t}{1-t} = 1 + \frac{2}{t-1} = 1 + \frac{2}{ (T_c^*-T)/\Delta T_c-2}.
\end{gather}

Since this expression does not depend on $x$, we conclude that far below $T^*_c$ (measured in units of $\Delta T_c$), the upper critical field anisotropy ratio has a temperature dependence which is independent of $J_4$ (See Main Text).

\begin{table}[t]
\centering
\begin{ruledtabular}
\begin{tabular}{ccl}
Symbol  & Definition & Physical meaning    \\ [4pt]
\hline
$l_b $ &   $1/l_b^2 = 2eH$ &  Magnetic length    \\ [4pt]
$\delta $ &   $g(u_{xx}-u_{yy})$ &  Strength symmetry breaking pinning field   \\ [4pt]
$\omega$ &   $\sqrt{J_1/J_3}$ &  Cyclotron frequency of the system  \\ [1pt]
& &  without crystal anisotropy effects    \\ [4pt]
$x$ &   $J_4 / J_1 $ & Dimensionless measure of $J_4$   \\ [4pt]
$y$ &  $\delta / \sqrt{J_1J_3} $ & Effective measure of symmetry breaking field  \\ [4pt]
$z$ &  $J_5 / \sqrt{J_1J_3} $ & Dimensionless measure of $J_5$  \\ [4pt]
$\Delta T_c$ & $|\delta|/A'$  & Shift of $T_c$ due to symmetry breaking field  \\ [4pt]
$T^*_c $ & $T_c+\Delta T_c$ &   The new critical temperature in the presence of symmetry breaking field   \\ [2pt]
$t $ & $- 1 - (T-T^*_c)/\Delta T_c$ &  Measure of temperature in units of $\Delta T_c$   \\ [4pt]
\end{tabular}
\end{ruledtabular}
 \caption{This Table lists and summarizes the definitions used in this section for calculating the upper critical field in the presence of a symmetry breaking field. }
\label{tab:defsSB}
\end{table}

Strictly speaking, Eq.~\eqref{eq:symbreak3} is valid when $J_4$ is small. We can also address a different limiting, in which no assumptions are made with regard to $J_4$, but instead we assume relatively large magnetic field (temperature is far below $T_c$), expressed as $eH\sqrt{J_3(J_1 - |J_4|)}\gg |\delta|$. In this case, the equation from which $H_{c2}$ can be determined reads as 
\begin{multline}
-\frac{A}{eH\sqrt{J_1 J_3}} = \sqrt{1-\frac{J_4}{J_1}} + \sqrt{1+ \frac{J_4}{J_1}}- \left[ 2-2\sqrt{1-\frac{J_4^2}{J_1^2}}+ \frac{2\delta}{eH\sqrt{J_1 J_3}}\cos2\theta \left(\sqrt{1-\frac{|J_4|}{J_1}} - \sqrt{1+ \frac{|J_4|}{J_1}}\right) \right. \\ \left. + \frac{\delta^2}{e^2 H^2 J_1 J_3}\left(   \cos^2 2\theta + \sin^2 2\theta \frac{2\sqrt[4]{1-\frac{J_4^2}{J_1^2}}}{\sqrt{1-\frac{J_4}{J_1}} + \sqrt{1+ \frac{J_4}{J_1}}} \right) \right]^{1/2}.
\end{multline}
This result is applicable as long as $|J_4|<J_1.$ If, in addition, we demand that $|\delta| \ll eH\sqrt{J_3}(\sqrt{J_1+|J_4|} - \sqrt{J_1-|J_4|}),$ the result takes the simple form
\be
H_{c2} = \frac{-A - \delta \, \text{sgn}(J_4) \cos 2\theta}{2e\sqrt{J_3 (J_1 - |J_4|)}} .
\ee

\subsubsection{Symmetry breaking field $\delta$ and $J_{4,5}$ }

As a next step we introduce the trigonal anisotropy term proportional to $J_5$. Including this term, the GL equations take the following form
\begin{gather}
-A  \eta = \left[ (J_1D^2_{\perp} + J_3 D^2_z)I + J_4D^2_{\perp} ( \cos 2\theta \tau^z+ \sin 2\theta \tau^x)+ J_5 \{D_z,D_{\perp}\} ( -\cos \theta \tau^z+ \sin \theta \tau^x) + \delta\tau^z \right] \eta  \label{eq:symbreakJ5}.
\end{gather}
In terms of the position and momentum operators the equations read
\begin{gather}
-\frac{A}{J_3}  \eta = \left[ \frac{1}{l^2_b}(\hat{P}^2 + \omega^2 \hat{X}^2 )I -  \frac{J_4}{J_3 l^2_b}\hat{X}^2 ( \cos 2\theta \tau^z+ \sin 2\theta \tau^x) +  \frac{J_5}{J_3 l^2_b} \{\hat{X},\hat{P}\} ( -\cos \theta \tau^z+ \sin \theta \tau^x) +  \frac{\delta}{J_3} \tau^z  \right] \eta  \label{eq:symbreakJ5_2}.
\end{gather}
Under the assumption that both $J_4$ and $J_5$ are small, we can construct a degenerate perturbation theory up to second order. Such perturbation theory up to first order is what produced Eq.~\eqref{eq:symbreak3}, and the effect of $J_5$ only enters at second order. Using the same basis states as in Eq.~\eqref{eq:basisstates1} and writing $\eta$ as $\eta = \alpha \ket{1} + \beta \ket{2}$, we find the followong matrix equation for the coefficients $\alpha,\beta$
\begin{gather}
-\frac{A}{J_3} \eta = \left[ \frac{\omega}{l^2_b}I -\frac{J^2_5}{2J^2_3 l^2_b\omega}I  - \frac{J^2_4}{8J^2_3 l^2_b\omega^3} I-  \frac{J_4}{2 J_3 l^2_b\omega} ( \cos 2\theta \tau^z+ \sin 2\theta \tau^x)    +  \frac{\delta}{J_3}\tau_z \right] \eta\label{eq:symbreakJ5_2}.
\end{gather}
In the same way as above, the upper critical is found by considering the lowest eigenvalue of the matrix equation and setting it equal to $-A/J_3\omega$. This straightforwardly gives the following generalization of Eq.~\eqref{eq:symbreak3}
\begin{eqnarray}
-\frac{A}{J_3\omega}  &=&  \frac{1}{l^2_b}\left(1 -\frac{J^2_5}{2J^2_3 \omega^2}  - \frac{J^2_4}{8J^2_3 \omega^4} \right) - \sqrt{ \frac{J^2_4}{4\omega^4 J^2_3 l^4_b} \sin^2 2\theta + \left(   \frac{\delta}{J_3\omega} - \frac{J_4}{2\omega^2 J_3 l^2_b}\cos 2\theta \right)^2  } \nonumber \\
 &=& \frac{1}{l^2_b}\left(1 -\frac{z^2}{2}  - \frac{x^2}{8} \right)  - \sqrt{ \frac{x^2}{4 l^4_b} \sin^2 2\theta + \left( \frac{z^2}{2 l^2_b }  - \frac{x^2}{8 l^2_b }   +  y - \frac{x}{2 l^2_b}\cos 2\theta \right)^2  } ,
\end{eqnarray}
where we have used the same definitions of $x,y$ as in Eq.~\eqref{eq:defxy}, in addition to the definition $z = J_5/J_3\omega = J_5/\sqrt{J_1 J_3}$. It is easy to see from this equation, that to second order in both $J_4$ and $J_5$ the corrections to Eq.~\eqref{eq:symbreak3} do not introduce qualitatively different behavior. 

One can carry out perturbation theory up to third order in both $J_4$ and $J_5$, which will introduce the sixfold anisotropy into the equation, entering as a term proportional to $\sim |J_4|J^2_5$. 

To conclude, we consider a final limiting case, in which we take $J_4=0$ and assume that we are in a regime of large fields, i.e., far below $T_c$ measured in units of $\delta$. In this case we must again resort to quasi-degenerate perturbation theory and derive the Hamiltonian in the low-energy subspace to second order in $J_5/J_3$. The calculation proceeds along the same line as the calculation based following Eq.~\eqref{eq:PTdegen1}. We find the Hamiltonian up to second order as
\begin{gather}
H^{(0)}_{ij} + H^{(2)}_{ij}  = \frac{\omega}{l^2_b}\left[1 - \frac{J^2_5}{2 \omega^2 J^2_3 }\left(\cos^2 \theta+\sin^2 \theta\frac{1}{1-(\overline{\delta}/2)^2} \right) + \overline{\delta}\tau^z - \frac{J^2_5\sin\theta}{2 \omega^2 J^2_3 } \frac{\overline{\delta}/2}{1-(\overline{\delta}/2)^2} \left(\sin \theta \tau^z+\cos \theta \tau^x \right)  \right], \label{eq:J5matrix}
\end{gather}
where we have defined $\overline{\delta} = l^2_b \delta/\omega$. Since $l^2_b \sim 1/H$ we see that this expression indeed only makes sense at high fields. Note that if we take symmetry breaking field to zero, $\delta \rightarrow 0$, we simply obtain
\begin{gather}
H^{(0)}_{ij} + H^{(2)}_{ij}  = \frac{\omega}{l^2_b}\left[1 - \frac{J^2_5}{2 \omega^2 J^2_3 } \right], 
\end{gather}
which is the correct result up to second order in $J_5/J_3$ according to Eq.~\eqref{eq:exactJ5only}. Diagonalizing the matrix of Eq.~\eqref{eq:J5matrix} and setting it equal to $-A/J_3$ yields the following implicit equation for the upper critical field 
\begin{gather}
-\frac{A}{J_3} = \frac{\omega}{l^2_b}\left[1 - \frac{J^2_5}{2 \omega^2 J^2_3 }\left(\cos^2 \theta+\sin^2 \theta\frac{1}{1-(\overline{\delta}/2)^2} \right) - \sqrt{\frac{J^4_5 \sin^2\theta}{4 \omega^4 J^4_3 }\frac{(\overline{\delta}/2)^2}{(1-\overline{\delta}^2/4)^2} + \overline{\delta}^2 - \overline{\delta} \frac{J^2_5\sin^2\theta}{\omega^2 J^2_3}  \frac{\overline{\delta}/2}{1-(\overline{\delta}/2)^2}  }   \right].
\end{gather}
From this equation we can already infer that for large fields, i.e., $\overline{\delta} \rightarrow 0 $ the twofold anisotropy of the angular dependence vanishes.

\subsection{The operator algebra of squeezed states\label{ssec:squeezing}}

In this section we provide additional information on the squeezing operators used to solve linearized GL equations in a magnetic field. The starting point is the simple harmonic oscillator Hamiltonian $\hat{P}^2 + \omega^2 \hat{X}^2$. The eigenvalues of this operator are found using the raising and lowering operators $\Pi_\pm  =  (\hat{P} + i \omega\hat{X})/\sqrt{2\omega}$, with $\Pi_+ = \Pi^\dagger_-$, which satisfy $[ \Pi_-, \Pi_+ ] =1$.

The squeezing operator is defined as
\begin{gather}
\hat{S}(z) = \exp\left\{ \frac{1}{2} (z^*\Pi^2_- - z \Pi^2_+ )  \right\}, \qquad z = r e^{i\theta}
\end{gather}
where $z$ is a complex number characterizing the squeezing. The name squeezing operator originates from the fact that momentum and position quadratures are ``squeezed'' under application of these unitary operators. Note that the squeezing operators are unitary and therefore preserve commutation relations.

Their action on the raising and lowering operators is given by
\begin{eqnarray}
\hat{S}^\dagger(z)\Pi_-  \hat{S}(z)  &= &  \cosh r \Pi_-  - e^{i\theta} \sinh r \Pi_+ \nonumber \\
\hat{S}^\dagger(z)\Pi_+  \hat{S}(z)  &= &  \cosh r \Pi_+  - e^{-i\theta} \sinh r \Pi_-
\end{eqnarray}
Since the position and momentum operators are related to $\Pi_+ \pm  \Pi_-$, we have the for these quantities after application of the squeezing operators
\begin{eqnarray}
\hat{S}^\dagger(z)\hat{P} \hat{S}(z)  &= & (\cosh r - \cos \theta \sinh r)\hat{P} + \omega \sin\theta\sinh r \hat{X} \nonumber \\
\hat{S}^\dagger(z)\hat{X} \hat{S}(z)  &= &  (\cosh r + \cos \theta \sinh r)\hat{X} +  \frac{1}{\omega}\sin\theta\sinh r \hat{P}
\end{eqnarray}
Note that in case $z$ is purely real, $\theta=0$, the momentum and position operators transform as
\begin{eqnarray}
\hat{S}^\dagger(z)\hat{P} \hat{S}(z)  &= & e^{-r}\hat{P} \nonumber \\
\hat{S}^\dagger(z)\hat{X} \hat{S}(z)  &= &  e^{r}\hat{X}
\end{eqnarray}
and it is precisely this rescaling which is a manifestation of squeezing. It is this property that will be used to express the solutions of two harmonic oscillator problems with different frequencies in terms of the same raising and lowering operators.

For the present purposes we need the matrix elements of the squeezing operators. In order to find them we first the structure of the squeezed vacuum. In particular, we relate the raising and lowering operation on the squeezed vacuum. We find
\begin{gather}
\hat{S}\Pi_- \ket{0} = 0  = \hat{S}\Pi_- \hat{S}^\dagger  \hat{S}\ket{0} =  ( \cosh r \Pi_-  +  \sinh r \Pi_+ ) \hat{S}\ket{0}, \quad \rightarrow \quad  \cosh r \Pi_-\hat{S}\ket{0} = - \sinh r \Pi_+  \hat{S}\ket{0}
\end{gather}
With the help of this relation we find the following expressions for the matrix elements that need to be calculated,
\begin{eqnarray}
\bra{n}\Pi_+\Pi_- \hat{S}\ket{0}  &=& n \bra{n}\hat{S}\ket{0}, \nonumber \\
\bra{n}\Pi_-\Pi_+ \hat{S}\ket{0}  &=& (n+1) \bra{n}\hat{S}\ket{0}, \nonumber \\
\bra{n}\Pi_+\Pi_+ \hat{S}\ket{0}  &=& -n \frac{\cosh r }{\sinh r}  \bra{n}\hat{S}\ket{0}, \nonumber \\
\bra{n}\Pi_-\Pi_- \hat{S}\ket{0}  &=& - (n+1 ) \frac{\sinh r }{\cosh r}\bra{n}\hat{S}\ket{0},
\end{eqnarray}
We find that the matrix elements $\bra{n}\hat{S}\ket{0}$ are the only objects we need. Furthermore, for our purposes we only require the absolute value squared $|\bra{n}\hat{S}\ket{0}|^2$, which is know is in the literature~\cite{albano02_SM}, and given by
\begin{gather}
|\bra{n}\hat{S}\ket{0}|^2 = \frac{n!}{(\tfrac{n}{2}!)^2 2^n}\frac{\tanh^n r}{\cosh r}, \quad n= 0,2,4,6,\ldots
\end{gather}

%
%
%
%

\section{Microscopic calculation of GL coefficients}

The purpose of this section is to calculate the GL coefficients, which up to this point have been treated as phenomenological expansion constants, from a microscopic model for paired electrons within a weak-coupling BCS approach. This will allow us to express the GL coefficients in terms of quantities characterizing the electronic normal state, i.e., the Fermi surface properties.

The calculation proceeds as follows. We first consider a microscopic mean-field theory in which pairs of electrons are coupled to a bosonic order parameter field. The mean-field action is quadratic in the fermionic fields, which can be integrated out to obtain an effective free energy the superconducting order parameter fields. The effective action is an expansion of this free energy in the order parameter fields (i.e., the Ginzburg-Landau free energy) with the expansion coefficients given by fermion loop diagrams. Explicit evaluation of the loop diagrams yields expressions for GL coefficients in terms of quantities directly related to the Fermi surface electrons.

The starting point of our calculation is a mean-field action $S = S_0 + S_\Delta$, expressed in terms of the electron operators $\psi_{\alpha}(\vec{x},\tau)$, where $S_\Delta$ represents a coupling to a pairing potential $ \hat{\Delta}(\vec{x},\vec{x}')$ (a matrix in spin space). Specifically, we start from the action
\begin{gather}
S = \int d\tau d^3\vec{x}\;  \psi^\dagger_{\alpha}(\vec{x},\tau) \Big[\partial_\tau + \varepsilon(-i\partial/\partial \vec x )- \ve_F \Big] \psi_{\alpha}(\vec{x},\tau) + \frac 12\int d\tau d^3\vec{x}d^3\vec{x}'\;  \Big[\hat \Delta_{\alpha\gamma}(\vec{x},\vec{x}')(i\sigma^y)_{\gamma\beta}\psi^\dagger_{\alpha}(\vec{x},\tau) \psi^\dagger_{\beta}(\vec{x}',\tau) + \text{h.c.}\Big]
\end{gather}
Here $ \varepsilon(\vec p )$ is the normal state kinetic energy as function of momentum operator $ \vec p= -i \partial / \partial \vec x $, and $\ve_F$ is Fermi energy. At this stage we do not include the electromagnetic gauge potential explicitly, but instead impose gauge invariance at the end.

Taking the Fourier transform, the superconducting part of the action is expressed as
\begin{gather}
S_\Delta =  \frac 12\int \frac{d\tau d^3\vec{k}d^3\vec{q}}{(2\pi)^6}\;  \hat \Delta_{\alpha\gamma}(\vec{q},\vec{k})(i\sigma^y)_{\gamma\beta}\psi^\dagger_{\alpha}(\vec{k}+\frac{\vec{q}}{2},\tau) \psi^\dagger_{\beta}(-\vec{k}+\frac{\vec{q}}{2},\tau) + \text{h.c.}, \label{eq:actiondelta}
\end{gather}
where $\vec{k}$ is the momentum variable conjugate to the relative coordinate, and $\vec{q}$ is conjugate to the center-of-mass coordinate as follows
\be
\hat \Delta(\vec q, \vec k) = \int d\vec x d\vec x' \hat \Delta (\vec x, \vec x') \exp[-i\vec q \cdot \frac{\vec x + \vec x'}2 - i\vec k \cdot (\vec x - \vec x') ].
 \ee
The superconducting order parameter $\hat{\Delta}(\vec{q},\vec{k}) $ is decomposed as $\hat{\Delta}(\vec{q},\vec{k}) = \sum_m \eta_m(\vec{q})\hat{\Delta}_m(\vec{k})$. Here $\eta_m(\vec{q})$ are the complex order parameters and $\hat{\Delta}_m(\vec{k})$ are the superconducting structure factors corresponding to component $m$ of the multi-dimensional representation [specified in Eq.~\eqref{eq:pairingeu} below].

To proceed, we define a Nambu spinor in particle-hole space as
\begin{gather}
\Phi(\vec{k},\tau)=  \begin{pmatrix} \psi_{\alpha}(\vec{k},\tau)  \\ \epsilon_{\alpha\beta} \psi^\dagger_{\beta}(-\vec{k},\tau)
\end{pmatrix}, \label{eq:nambu}
\end{gather}
and with this definition the integrand of~\eqref{eq:actiondelta} can be expressed as
\begin{gather}
\Phi^\dagger (\vec{k}+\frac{\vec{q}}{2},\tau) \hat{\Sigma}_{\Delta}(\vec{k},\vec{q}) \Phi(\vec{k}-\frac{\vec{q}}{2},\tau),
\end{gather}
where the mean-field superconducting self energy $ \hat{\Sigma}_{\Delta}$ is given by
\begin{gather}
\hat{\Sigma}_{\Delta}(\vec{k},\vec{q}) =  \begin{pmatrix}  0 & {\hat \Delta}(\vec{q},\vec{k}) \\ {\hat \Delta}^\dagger(-\vec{q},\vec{k}) &  0 \end{pmatrix} = {\hat \Delta}(\vec{q},\vec{k}) \tau^+ +{\hat \Delta}^\dagger(-\vec{q},\vec{k}) \tau^- \label{eq:selfdelta}
\end{gather}
with $\tau^{\pm} = (1/2)(\tau^x \pm i \tau^y)$.

We are now in the position to write down the full action $S = S_0 + S_{\Delta} $, which is bilinear in the electron operators and can be integrated to obtain an effective $S_{\text{eff}}[\eta_i]$. First, we transform to Matsubara frequency space and write $k  = ( i\omega, \vec{k})$ and $q = (0,\vec{q})$. Note that we have set the bosonic frequencies to zero since we will only be interested in static spatial inhomogeneities. The full electronic action $S$ then takes the form
\begin{gather}
S = -\frac12\sum_{p,q} \Phi^\dagger (k) \hat{\mathcal{G}}^{-1}_0(k) \Phi(k) + \frac12 \sum_{p,q} \Phi^\dagger (k+\frac{q}{2}) \hat{\Sigma}_{\Delta}(\vec{k},\vec{q}) \Phi(k-\frac{q}{2}),
\end{gather}
where we omitted a constant part that does not contain $\Phi$ operators. Here $\hat{\mathcal{G}}^{-1}_0(k)$ is the inverse Fermi liquid Green's function, $\hat{\mathcal{G}}_0(\vec k, \tau) = -\langle \textbf{T}_{\tau} \Phi_{\vec k}(\tau) \Phi_{\vec k}^\dagger(0)\rangle$,  given by
\begin{gather}
\hat{\mathcal{G}}^{-1}_0( k) = i\omega - \xi(\vec{k})\tau^z.
\end{gather}
We have abbreviated $\xi(\vec{k}) = \varepsilon(\vec{k})-\ve_F $. As a result, the Green's function $\hat{\mathcal{G}}_0$ is block diagonal, with the electron and hole Green's functions $G_\pm$ on the diagonal. The hole Green's function is obtained from particle-hole conjugation of the electron Green's function, and one has
\begin{gather}
\hat{\mathcal{G}}_0 = \begin{pmatrix}\hat{G}_+ &  \\
   & \hat{G}_- \end{pmatrix}= \begin{pmatrix} \hat{G}_+(\vec{k},i\omega)  &
   \\    & - \hat{G}_+(\vec{k},-i\omega)
 \end{pmatrix}.
\end{gather}
The effective action $S_{\text{eff}}[\eta_i] = (1/T) F[\eta_i] $ is obtained by the standard procedure of integrating out the fermionic degrees of freedom,
\begin{gather}
e^{-S_{\text{eff}}[\eta_i]}  = \int \mathcal{D}\psi^\dagger\mathcal{D}\psi \;e^{-S[\eta_i,\psi^\dagger,\psi]}.
\end{gather}
The quasiparticle part of the mean-field free energy can be expressed as
\begin{gather}
F[\eta_i] = - T \, \text{Tr}\, \ln\hat{\mathcal{G}}^{-1} = F_0
- T\, \text{Tr}\, \ln ( 1- \hat{\mathcal{G}}_0\hat{\Sigma}_{\Delta}), \label{eq:trace}
\end{gather}
where $F_0$ is the normal state part of free energy. Here, the trace $\text{Tr}$ is understood as a sum over frequency and momenta, and as a matrix trace over the matrix structure of $\hat{\mathcal{G}}_0$ and $\hat{\Sigma}_{\Delta}$: $\text{Tr} \equiv \sum_\omega \sum_{\vec{k}} \text{tr} $. We note that since we restrict to the quasiparticle part of the free energy, we have left the contribution to the free energy which is quadratic in the order parameter field and explicitly depends on the pairing interaction implicit.

We now focus on the derivation of the free energy to the second order in $\hat \Delta$. Explicitly, it is given by the expression
\begin{gather}
F[\eta_i] = F_0 + \frac{T}{2} \trace{\hat{\mathcal{G}}_0\hat{\Sigma}\hat{\mathcal{G}}_0\hat{\Sigma}  }  = F_0 + \frac{T}{2} \sum_{\omega}\sum_{\vec{k},\vec{q}}\text{tr}\; [\hat{\mathcal{G}}_0(\vec{k}+\frac{\vec{q}}{2},i\omega)\hat{\Sigma}_\Delta(\vec{q},\vec{k})\hat{\mathcal{G}}_0(\vec{k}-\frac{\vec{q}}{2},i\omega)\hat{\Sigma}_\Delta(-\vec{q},\vec{k}) ].
\end{gather}
The self-energy $\hat{\Sigma}_{\Delta}$ contains the order parameter fields and using Eq.~\eqref{eq:selfdelta} the expression for the superconducting part of free energy can be rewritten in the form
\begin{gather}
 F_{\Delta} = \frac{T}{2} \trace{\hat{\mathcal{G}}_0\hat{\Sigma}\hat{\mathcal{G}}_0\hat{\Sigma}  }  =  \sum_{\vec{q}}\sum_{m,n} \eta_m^*(\vec{q})\mathcal{Q}_{mn}(\vec{q})\eta_n(\vec{q}),
\end{gather}
where $\eta_n(\vec{q})$ are the order parameter fields and the matrix $\mathcal{Q}_{mn}(\vec{q})$ is given by
\begin{eqnarray}
\mathcal{Q}_{mn}(\vec{q}) &=&  T \sum_{\omega}\sum_{\vec{k}}G_+(\vec{k}+\frac{\vec{q}}{2},i\omega)G_-(\vec{k}-\frac{\vec{q}}{2},i\omega) \text{tr}\; [{\hat \Delta}_n(\vec{k}){\hat \Delta}^\dagger_m(\vec{k}) ] \nonumber \\
& \equiv&  T \sum_{\omega}\sum_{\vec{k}}G_+(\vec{k}+\frac{\vec{q}}{2},i\omega)G_-(\vec{k}-\frac{\vec{q}}{2},i\omega) Q_{mn}(\vec{k}).
\end{eqnarray}
Here we have defined the form factor matrix $Q_{mn}(\vec{k}) =  \text{tr}\; [{\hat \Delta}_n(\vec{k}){\hat \Delta}^\dagger_m(\vec{k}) ] $.

Expanding now the electron Green functions in small momenta $\vec q$ using
\begin{gather}
\xi(\vec{k}\pm \frac{\vec{q}}{2}) = \xi(\vec{k}) \pm \frac{1}{2}\vec{v}(\vec{k})\cdot \vec{q} + \mathcal{O}(q^2),
\end{gather}
(the second order term turns out to make parametrically smaller contribution) we obtain eventually
\be
\mathcal{Q}_{mn}(\vec q) \approx - T \sum_{\omega, \vec k} \frac1{\omega^2 + \xi^2(\vec k)}\left(  1 + \frac{[\vec v (\vec k)\cdot \vec q]^2}4 \frac{\xi^2(\vec k) - 3\omega^2}{[\omega^2 + \xi^2(\vec k)]^2} \right)Q_{mn}(\vec k). \label{eq: Q}
\ee
This expression is the starting point for the microscopic derivation of the coefficient in our GL theory.

Below we will consider small distortions of the Fermi surface, either due to trigonal crystal anisotropy, or uniaxial strain-induced anisotropy. In this case, one can write $\xi(\vec k) = \xi_0(\vec k) + \delta \xi (\vec k)$, $\vec v(\vec k) = \vec v_0 (\vec k) + \delta \vec v (\vec k),$ with $\xi_0(\vec k) = k^2/2m - \ve_F$, $\vec v_0(\vec k) = \vec k/m$ and $\delta \vec v(\vec k) = \partial \delta \xi (\vec k) / \partial \vec k$. To the leading order, gradient term can be expanded as
\be
\frac{(\vec v \cdot \vec q)^2(\xi^2-3\omega^2)}{(\xi^2+\omega^2)^3}\approx \frac{(\vec v_0 \cdot \vec q)^2 (\xi_0^2-3\omega^2)}{(\omega^2+\xi_0^2)^3} + 2\frac{(\vec v_0 \cdot \vec q) ( \delta \vec v \cdot \vec q) (\xi_0^2-3\omega^2)}{(\omega^2 + \xi_0^2)^3} + 4\xi_0 \frac{(\vec v_0 \cdot \vec q)^2 (5\omega^2 - \xi_0^2)}{(\omega^2 + \xi_0^2)^4}\delta \xi. \label{eq: expansion}
\ee

\subsection{Calculation of gradient coefficients $J_{1,2,3,4,5}$ in presence of trigonal crystal anisotropy}

We now proceed to the calculation of the gradient coefficients $J_{1,2,3,4,5}$. In particular, we demonstrate that the trigonal gradient term $f_{D,\text{trig}}$ (see Main Text) is generated only when the Fermi surface has trigonal crystal anisotropy, or when the gap functions are general linear combinations of crystal harmonics allowed by trigonal symmetry. Hence, we take into account the trigonal anisotropy of the Fermi surface (FS) and the specific form of the pairing potential $\hat \Delta_m(\vec k)$ applicable to trigonal symmetry.

In the case of trigonal distortion, energy spectrum of electrons can be described as
\be
\xi(\vec k) = \xi_0(\vec k) + \delta \xi(\vec k) = \xi_0(\vec k) + i \frac{\lambda_{\text{trig}}}{2mk_F^2} k_z (k_+^3 - k_-^3)=\xi_0(\vec k) + \frac{\lambda_{\text{trig}}}{m k_F^2}k_z k_y(k_y^2 - 3k_x^2),
\ee
where $\xi_0=k^2/2m - \ve_F$, $k_{\pm} = k_x \pm i k_y$ and $k_F=\sqrt{2m\ve_F}$ is Fermi momentum. Here we have neglected all other fourth order contribution to the dispersion, since they affect the result only quantitatively and in a way unimportant for our purposes. In particular, they do not affect the gradient coefficient of $f_{D,\text{trig}}$, which is what we are mainly interested in. The corresponding change in the velocity equals
\be
\delta \vec v(\vec k) = \frac{\lambda_{\text{trig}}}{2mk_F^2} \left( \begin{array}{c} -6 k_x k_y k_z \\ 3k_z k_y^2 - 3 k_z k_x^2 \\ k_y^3 - 3 k_x^2 k_y  \end{array}   \right).
\ee

Then, the recipe is to plug $\delta \xi$ and $\delta \vec v$ back into Eq. (\ref{eq: expansion}) and perform the integration over $\xi_0$, the summation over $\omega$ and the average over the directions of $\vec k$. Though straightforward, this procedure is somewhat tedious. We note that the third term in (\ref{eq: expansion}) is odd in $\xi_0$ and thus, to the leading order, is zero after integration. To obtain a non-zero contribution, one needs to carefully extract the $\xi_0$ dependence, i.e., take $k\approx k_F(1+\xi_0/2\ve_F)$ everywhere. In addition, the $\xi_0$ dependence of density of states must be taken into account. In three dimensions, and for a quadratic spectrum, it equals $N(\ve_F+\xi_0) \approx N(\ve_F)(1+\xi_0/2\ve_F).$

In order to perform the averaging over angles, we calculate form factors $Q_{mn}(\vec{k}) =  \text{tr}\; [{\hat \Delta}_n(\vec{k}){\hat \Delta}^\dagger_m(\vec{k}) ] $. For the odd-parity two-component pairing in crystals with trigonal symmetry ($D_{3d}$) pairing potentials $\hat \Delta_{1,2}(\vec k)$ are given by
\begin{gather}
\Delta_1(\vec{k}) = \lambda_a \hat{k}_x\sigma^z + \lambda_b\hat{k}_z\sigma^x + \lambda_c(\hat{k}_y\sigma^x + \hat{k}_x\sigma^y), \quad
 \Delta_2(\vec{k}) =\lambda_a\hat{k}_y\sigma^z +  \lambda_b\hat{k}_z\sigma^y + \lambda_c(\hat{k}_x\sigma^x - \hat{k}_y\sigma^y) \label{eq:pairingeu}
\end{gather}
with real coefficients $\lambda_{a,b,c}.$ In hexagonal crystals, symmetry imposes the constraint $\lambda_c=0$ ($E_{1u}$ pairing) or $\lambda_a = \lambda_b = 0$ ($E_{2u}$ pairing), whereas in trigonal crystals these p-wave spherical harmonics have $E_{u}$ symmetry. With this pairing, we find for form-factors
\begin{eqnarray}
Q_{11}  &=  &2\lambda^2_a \hat{k}^2_x +  2\lambda^2_b \hat{k}^2_z + 2\lambda^2_c( \hat{k}^2_x +\hat{k}^2_y ) +4\lambda_b\lambda_c\hat{k}_z\hat{k}_y, \nonumber \\
Q_{22}  &=  &2\lambda^2_a \hat{k}^2_y +  2\lambda^2_b \hat{k}^2_z + 2\lambda^2_c( \hat{k}^2_x +\hat{k}^2_y ) -4\lambda_b\lambda_c\hat{k}_z\hat{k}_y, \nonumber \\
Q_{12}  = Q_{21} &=  & 2 \lambda^2_a \hat{k}_x\hat{k}_y + 4\lambda_b\lambda_c\hat{k}_z\hat{k}_x.
\end{eqnarray}

After the straightforward calculations, the expression for the gradient part of free energy reads as (we omit indices $m, n$ for brevity)
\be
F_{\nabla} = \sum_{\vec q} \eta^\dagger(\vec q) \mathcal{Q}(\vec q) \eta(\vec q)
\ee
where
\be
\mathcal{Q}(\vec q) = J_1 (q^2_x + q^2_y)I+ J_3q^2_z I  +  J_4 \left[(q^2_x - q^2_y)\tau^z +2q_xq_y\tau^x  \right]+  2 J_5 \left[ q_zq_y\tau^z +q_zq_x\tau^x  \right].
\ee
Coefficients $J_i$ are given by
\begin{gather}
J_i= \tilde{J}_i \frac{7 \zeta(3) N(\ve_F) v^2_F}{120\pi^2 T^2}.
\end{gather}

\be
\tilde{J}_1 = 2\lambda^2_a + \lambda^2_b+4 \lambda^2_c, \quad \tilde{J}_3 = \lambda^2_a + 3\lambda^2_b + 2\lambda^2_c , \quad \tilde{J}_4 = \lambda^2_a- 2c \lambda_{\text{trig}}\lambda_b\lambda_c, \quad \tilde{J}_5 = 2 \lambda_b \lambda_c -c\lambda_{\text{trig}}\lambda_a^2,
\ee
with $c=22/21$.

We see that, indeed, trigonal anisotropy of Fermi surface and special form of the pairing function lead to the generation of the trigonal gradient term $J_5$. We see also that, to the leading order, the term $J_2$ is absent. It becomes non-zero if one takes into account particle-hole unsymmetry (dependence of density of states on energy). This term is of the order $(T_c/\ve_F)^2\ll1.$

\subsection{Calculation of contributions of the symmetry breaking field}

The effect of the symmetry breaking field can be described by the Hamiltonian
\be
H_{\text{SB}} =  \frac{\lambda_{\text{SB}}}{2m} \sum_{\vec{k}} \psi^\dagger(\vec k) \psi(\vec k)(k_x^2-k_y^2).
\ee
For simplicity, we particularize to the case of hexagonal symmetry [i.e. take $\lambda_c=0$ in (\ref{eq:pairingeu})] and do not take into account any fourth-order terms in dispersion relation. The presence of the symmetry breaking field leads to the dispersion relation
\be
\xi(\vec k) = \frac{k^2}{2m} - \ve_F + \frac{\lambda_{\text{SB}}}{2m}(k_x^2-k_y^2),
\ee
with $\delta \xi(\vec k) = \lambda_{\text{SB}}(k_x^2 - k_y^2)/2m.$ Again, to extract the key physical features we treat $\lambda_{\text{SB}}$ as a small perturbation (in principle, the problem can be solved exactly for any finite $\lambda_{\text{SB}}$).

To the leading order, the symmetry-breaking field couples to the superconducting order parameter, according to Eq.~\eqref{eq:mncouple}. To find this coupling explicitly, we consider the first term in Eq. (\ref{eq: Q}), and write (we omit indices $m,n$)
\be
\mathcal Q_{0}^{\text{SB}} = 2 T \sum_{\omega,\vec k} \frac{\xi_0(\vec k) \delta \xi(\vec k)}{[\omega^2 + \xi_0^2(\vec k)]^2} Q(\vec k).
\ee
Again, we carefully extract $\xi_0$ to obtain the leading non-vanishing contribution.

After the integration over $\xi_0$, we end up with the formally diverging sum over $\omega$, which requires a cut-off regularization by the (Debye) frequency $\omega_{\max} \sim \omega_{D}$:
\be
\sum_{\omega} \int d\xi_0 \frac{\xi_0^2}{(\omega^2+\xi_0^2)^2} \approx \frac1{2T}\ln \frac{\omega_D}T.
\ee
After averaging over directions of $\vec k$, we find
\be
\mathcal Q_{0}^{\text{SB}} \approx \frac{2}{5} \lambda_{\text{SB}}  N(\ve_F) \lambda_a^2 \ln \frac{\omega_D}{T} \tau^z,
\ee
or, equivalently, for free energy
\be
F^{\text{SB}}_0 = \sum_{\vec q}\eta^\dagger(\vec q) \mathcal Q_{0}^{\text{SB}}\eta (\vec q) = \left(\frac{2}{5} \lambda_{\text{SB}}  N(\ve_F) \lambda_a^2 \ln \frac{\omega_D}{T}  \right) \sum_{\vec q} |\eta_1(\vec q)|^2 - |\eta_2(\vec q)|^2 , \label{eq:Flambda}
\ee
This term is responsible for the shift of $T_c$, $\Delta T_c / T_c \sim \lambda_{\text{SB}} \ln\omega_D/T$. This implies that the effect of a strain-induced Fermi surface distortion ($\lambda_{\text{SB}}$) on the shift of $T_c$ is enhanced by $ \ln\omega_D/T$. The coefficient on the right hand side of \eqref{eq:Flambda} is what we have called $\delta$ in the previous sections and Main Text. 

Next, we calculate the gradient terms due to symmetry-breaking field, i.e., the second term in Eq. (\ref{eq: Q}). Velocity is given now by
\be
\delta \vec v(\vec k) = \frac{\lambda_{\text{SB}}}m\left( \begin{array}{c} k_x \\ -k_y \\ 0      \end{array}  \right).
\ee
After straightforward integration, we find
\be
F^{\text{SB}}_{\nabla} = \sum_{\vec q}\eta^\dagger(\vec q) \mathcal Q_{\nabla}^{\text{SB}}(\vec q) \eta (\vec q),
\ee
with the momentum-dependent matrix $\mathcal Q_{\nabla}^{\text{SB}}(\vec q)$ given by
\be
\mathcal Q_{\nabla}^{\text{SB}}(\vec q) = K_1 (q_x^2 - q_y^2) I + K_2 (q_x^2 + q_y^2) \tau^z + K_3 q_z^2 \tau^z.
\ee
These strain-induced contributions to the gradient terms have gradient coefficients given by
\be
K_i = \tilde K_i \frac{\zeta(3)N(\ve_F) v_F^2}{120\pi^2T^2} \lambda_{\text{SB}},
\ee
where the $\tilde K_i$ are given by
\be
\tilde K_1 = 13\lambda_a^2 + 9 \lambda_b^2, \qquad \tilde K_2 = -\lambda_a^2, \qquad \tilde K_3 = -5 \lambda_a^2.
\ee

It is instructive to compare the coupling of symmetry-breaking field to the order-parameter and its derivatives. It is convenient to introduce the coherence length $\xi= \xi(T) $ and $\xi_{0} = \xi(T=0) \sim v_F/T_c$ (Note that here $\xi$ is used for coherence length). The ratio of the gradient term to the direct quadratic coupling to the order parameter equals
\be
\frac{F^{SB}_{\nabla}}{F^{SB}_{0}} \sim \frac{v_F^2 q^2}{T^2 \ln \omega_D/T} \sim \frac{(\xi_{0}q)^2}{\ln \omega_D/T}.
\ee
The relevant momenta are those where $q\sim 1/\xi$. Close to the transition temperature, $\xi_{0}\ll \xi$, and we have
\be
\frac{F^{SB}_{\nabla}}{F^{SB}_{0}} \sim \left(  \frac{\xi_{0}}{\xi} \right)^2 \frac1{\ln \omega_D/T} \ll 1.
\ee
It follows from our calculation that the direct coupling of symmetry-breaking field to order parameter is much stronger than to its derivatives. It implies that the effect of uniaxial distortion field is much more pronounced in the case of two-component superconductors. Indeed, the effect of the symmetry breaking field in case of two-component order parameters is to shift the transition temperature.  Single-component superconductors allow coupling to the derivatives of the order parameter only, thus significantly decreasing possible effects of the symmetry-breaking field.


\bibliographystyle{apsrev}

\begin{thebibliography}{10}


\bibitem{sigrist91} M. Sigrist and K. Ueda, Rev. Mod. Phys. {\bf 63}, 239 (1991).

\bibitem{mineevbook} V. Mineev and K. Samokhin,  {\it Introduction to Unconventional Superconducitivity}, (Gordon and Breach, New York, 1999).

\bibitem{harlingen95} D. J. Van Harlingen, Rev. Mod. Phys. {\bf 67}, 515 (1995).

\bibitem{tsuei00} C. C. Tsuei and J. R. Kirtley, Rev. Mod. Phys. {\bf 72}, 969 (2000).

\bibitem{mackenzie03} A. Mackenzie and Y. Maeno, Rev. Mod. Phys. {\bf 75}, 657 (2003). 

\bibitem{volovik85} G. E. Volovik and L. Gork'ov, Sov. Phys. JETP {\bf 61}, 843 (1985).

\bibitem{cava} Y. S. Hor, A. J. Williams, J. G. Checkelsky, P. Roushan, J. Seo, Q. Xu, H. W. Zandbergen, A. Yazdani, N. P. Ong, R. J. Cava, Phys. Rev. Lett. {\bf 104}, 057001 (2010).

\bibitem{ando} M. Kriener, Kouji Segawa, Zhi Ren, Satoshi Sasaki, and Yoichi Ando, Phys. Rev. Lett. {\bf 106}, 127004 (2011). 

\bibitem{fuberg} L. Fu and E. Berg, Phys. Rev. Lett. {\bf 105}, 097001 (2010).

\bibitem{hao11} L. Hao and T. K. Lee, Phys. Rev. B {\bf 83}, 134516 (2011).

\bibitem{bay12} T. V. Bay, T. Naka, Y. K. Huang, H. Luigjes, M. S. Golden, and A. de Visser, Phys. Rev. Lett. { \bf 108}, 057001 (2012).

\bibitem{nagai12} Y. Nagai, H. Nakamura, and M. Machida, Phys. Rev. B {\bf 86}, 094507 (2012).

\bibitem{lawson12} B. J. Lawson, Y. S. Hor, and L. Li, Phys. Rev. Lett. {\bf 109}, 226406 (2012)

\bibitem{hashimoto13} T. Hashimoto, K. Yada, A. Yamakage, M. Sato, and Y. Tanaka, J. Phys. Soc. Jpn. {\bf 82}, 044704 (2013)

\bibitem{yip13} S.-K. Yip, Phys. Rev. B {\bf 87}, 104505 (2013).

\bibitem{wan14} X. Wan and S. Y. Savrasov, Nat. Comm. {\bf 5}, 4144 (2014).

\bibitem{brydon14} P. M. R. Brydon, S. Das Sarma, H.-Y. Hui, J. D. Sau, Phys. Rev. B {\bf 90}, 184512 (2014).

\bibitem{schneeloch15} J. A. Schneeloch, R. D. Zhong, Z. J. Xu, G. D. Gu, and J. M. Tranquada, Phys. Rev. B {\bf 91}, 144506 (2015).

\bibitem{andofu} Y. Ando and L. Fu, Ann. Rev. Condens. Mater. Phys. {\bf 6}, 361 (2015).

\bibitem{point-contact} S. Sasaki, M. Kriener, K. Segawa, K. Yada, Y. Tanaka, M. Sato, and Y. Ando, Phys. Rev. Lett.  {\bf 107}, 217001 (2011).

\bibitem{stm} N. Levy, T. Zhang, J. Ha, F. Sharifi, A. A. Talin, Y. Kuk, and J. A. Stroscio, Phys. Rev. Lett.  {\bf 110}, 117001 (2013).

\bibitem{Zheng} K. Matano,  M. Kriener, K. Segawa, Y. Ando, and Guo-qing Zheng, arXiv:1512.07086 (2015). 

\bibitem{maeno} S. Yonezawa, K. Tajiri, S. Nakata, Y. Nagai, Z. Wang, K. Segawa, Y. Ando, and Y. Maeno, arXiv:1602.08941 (2016). 

\bibitem{fu14} Liang Fu, Phys. Rev. B {\bf 90}, 100509(R) (2015).


\bibitem{kivelson98} S. A. Kivelson, E. Fradkin, and V. J. Emery, Nature {\bf 393}, 550-553 (1998). 

\bibitem{fradkin} E. Fradkin, S. A. Kivelson, M. J. Lawler, J. P. Eisenstein, A. P. Mackenzie, Annu. Rev. Condens. Matter Phys. {\bf 1}, 153 (2010).

\bibitem{mineev} P. L. Krotkov, and V. P. Mineev, Phys. Rev. B {\bf 65}, 224506 (2002).

\bibitem{machida12} Y. Machida, A. Itoh, Y. So, K. Izawa, Y. Haga, E. Yamamoto, N. Kimura, Y. Onuki, Y. Tsutsumi, and K. Machida, Phys. Rev. Lett. {\bf 108}, 157002 (2012).

\bibitem{joynt02}  R. Joynt and L. Taillefer, Rev. Mod. Phys. {\bf 74}, 235 (2002). 

\bibitem{fisher89} R. A. Fisher, S. Kim, B. F. Woodfield, N. E. Phillips, L. Taillefer, K. Hasselbach, J. Flouquet, A. L. Giorgi, and J. L. Smith, Phys. Rev. Lett. {\bf 62}, 1411 (1989).

\bibitem{aeppli88} G. Aeppli, E. Bucher, C. Broholm, J. K. Kjems, J. Baumann, and J. Hufnagl, Phys. Rev. Lett. {\bf 60}, 615 (1988).

\bibitem{sr-doped} Zhongheng Liu, Xiong Yao, Jifeng Shao, Ming Zuo, Li Pi, Shun Tan, Changjin Zhang, Yuheng Zhang, J. Am. Chem. Soc. {\bf 137}, 10512 (2015).

\bibitem{sr-doped-2} Shruti, V. K. Maurya, P. Neha, P. Srivastava, and S. Patnaik, Phys. Rev. B {\bf 92}, 020506(R) (2015). 

\bibitem{nb-doped} Y. Qiu, K. Nocona Sanders, J. Dai, J. E. Medvedeva, W. Wu, P. Ghaemi, T. Vojta, Y. S. Hor, arXiv:1512.03519 (2015).

\bibitem{tl-doped} Z. Wang, A. A. Taskin, T. Fr\"olich, M Braden, and Y. Ando, Chem. Mater. {\bf 28}, 779 (2016).

\bibitem{Venderbos} J. W. F. Venderbos, V. Kozii, L. Fu, arXiv: arXiv:1512.04554 (2015).

\bibitem{hicks14} C. W. Hicks, D. O. Brodsky, E. A. Yelland, A. S. Gibbs, J. A. N. Bruin, M. E. Barber, S. D. Edkins, K. Nishimura, S. Yonezawa, Y. Maeno, A. P. Mackenzie, Science {\bf 344}, 283 (2014).

\bibitem{sauls94} J. A. Sauls, Adv. Phys. {\bf 43}, 113 (1994). 

\bibitem{zhitomirsky95} I. A. Luk'yanchuk, M .E. Zhitomirsky, Superconductivity Review 1, 207 (1995). 




\bibitem{suppmat} See Supplemental Material at ... .

\bibitem{gorkov84} L. P. Gor'kov, Zh. Eksp. Teor. Fiz. {\bf 40}, 351 (1984) [JETP Lett. {\bf 40}, 1155 (1984)].

\bibitem{burlachkov85} L. I. Burlachkov, Zh. Eksp. Teor. Fiz. {\bf 89}, 1382 (1985)  [Sov. Phys. JETP {\bf 62}, 800 (1985)].

\bibitem{machida85} K. Machida, T. Ohmi, and M. Ozaki, J. Phys. Soc. Jpn. {\bf 54}, 1552 (1985).

\bibitem{hess89} D. Hess, T. Tokuyasu, and J. A. Sauls, J. Phys. Condens. Matter {\bf 1}, 8135 (1989)

\bibitem{machida89} K. Machida and M. Ozaki, J. Phys. Soc. Jpn. {\bf 58}, 2244 (1989).

\bibitem{agterberg94} D. F. Agterberg and M. B. Walker, Phys. Rev. B {\bf 51}, 8481 (1994).

\bibitem{agterberg95} D. F. Agterberg and M. B. Walker, Phys. Rev. Lett. {\bf 74}, 3904 (1995).

\bibitem{sauls96} J. A. Sauls, Phys. Rev. B {\bf 53}, 8543 (1996).






\end{thebibliography}

\begin{thebibliography}{10}

\bibitem{volovik85_SM} G. E. Volovik and L. Gork'ov, Sov. Phys. JETP {\bf 61}, 843 (1985).

\bibitem{sigrist91_SM} M. Sigrist and K. Ueda, Rev. Mod. Phys. {\bf 63}, 239 (1991).

\bibitem{mineevbook_SM} V. Mineev and K. Samokhin,  {\it Introduction to Unconventional Superconducitivity}, (Gordon and Breach, New York, 1999).


\bibitem{sauls94_SM} J. A. Sauls, Adv. Phys. {\bf 43}, 113 (1994). 

\bibitem{zhitomirsky95_SM} I. A. Luk'yanchuk, M .E. Zhitomirsky, Superconductivity Review 1, 207 (1995). 

\bibitem{zhitomirsky90_SM} M .E. Zhitomirsky, Zh. Eksp. Teor. Fiz. {\bf 97}, 1346 (1990) [Sov. Phys. JETP {\bf 70}, 760-768 (1990)].

\bibitem{barash91_SM} Yu. S. Barash and A. V. Galaktionov, Zh. Eksp. Teor. Fiz. {\bf 100}, 1699 (1991) [Sov. Phys. JETP {\bf 73}, 939 (1991)]. 

\bibitem{mineev_SM} P. L. Krotkov, and V. P. Mineev, Phys. Rev. B {\bf 65}, 224506 (2002).


\bibitem{gorkov84_SM} L. P. Gor'kov, Zh. Eksp. Teor. Fiz. {\bf 40}, 351 (1984) [JETP Lett. {\bf 40}, 1155 (1984)].

\bibitem{burlachkov85_SM} L. I. Burlachkov, Zh. Eksp. Teor. Fiz. {\bf 89}, 1382 (1985)  [Sov. Phys. JETP {\bf 62}, 800 (1985)].

\bibitem{machida85_SM} K. Machida, T. Ohmi, and M. Ozaki, J. Phys. Soc. Jpn. {\bf 54}, 1552 (1985).

\bibitem{albano02_SM}  L. Albano, D. F. Mundarain, J. Stephany J. Opt. B Quant. Semiclass. Opt. {\bf 4}, 352 (2002).


\bibitem{agterberg94_SM} D. F. Agterberg and M. B. Walker, Phys. Rev. B {\bf 51}, 8481 (1994).

\bibitem{agterberg95_SM} D. F. Agterberg and M. B. Walker, Phys. Rev. Lett. {\bf 74}, 3904 (1995).










\end{thebibliography}

\end{document}